\documentclass[11pt]{article}
\usepackage{latexsym}
\def\AFOUR{%
\setlength{\textheight}{8.5in}%
\setlength{\textwidth}{5.75in}%
\setlength{\topmargin}{-0.375in}%
\hoffset=-.5in%
\renewcommand{\baselinestretch}{1.17}%
\setlength{\parskip}{6pt plus 2pt}%
}

\AFOUR                                           %%%   ***

\expandafter\ifx\csname amssym.def\endcsname\relax \else\endinput\fi
\expandafter\edef\csname amssym.def\endcsname{%
       \catcode`\noexpand\@=\the\catcode`\@\space}
\catcode`\@=11
\def\undefine#1{\let#1\undefined}
\def\newsymbol#1#2#3#4#5{\let\next@\relax
 \ifnum#2=\@ne\let\next@\msafam@\else
 \ifnum#2=\tw@\let\next@\msbfam@\fi\fi
 \mathchardef#1="#3\next@#4#5}
\def\mathhexbox@#1#2#3{\relax
 \ifmmode\mathpalette{}{\m@th\mathchar"#1#2#3}%
 \else\leavevmode\hbox{$\m@th\mathchar"#1#2#3$}\fi}
\def\hexnumber@#1{\ifcase#1 0\or 1\or 2\or 3\or 4\or 5\or 6\or 7\or 8\or
 9\or A\or B\or C\or D\or E\or F\fi}

\font\tenmsa=msam10
\font\sevenmsa=msam7
\font\fivemsa=msam5
\newfam\msafam
\textfont\msafam=\tenmsa
\scriptfont\msafam=\sevenmsa
\scriptscriptfont\msafam=\fivemsa
\edef\msafam@{\hexnumber@\msafam}
\mathchardef\dabar@"0\msafam@39
\def\dashrightarrow{\mathrel{\dabar@\dabar@\mathchar"0\msafam@4B}}
\def\dashleftarrow{\mathrel{\mathchar"0\msafam@4C\dabar@\dabar@}}

\def\ulcorner{\delimiter"4\msafam@70\msafam@70 }
\def\urcorner{\delimiter"5\msafam@71\msafam@71 }
\def\llcorner{\delimiter"4\msafam@78\msafam@78 }
\def\lrcorner{\delimiter"5\msafam@79\msafam@79 }
\def\yen{{\mathhexbox@\msafam@55}}
\def\checkmark{{\mathhexbox@\msafam@58}}
\def\circledR{{\mathhexbox@\msafam@72}}
\def\maltese{{\mathhexbox@\msafam@7A}}
\def\circledS{{\mathhexbox@\msafam@73}}

\font\tenmsb=msbm10
\font\sevenmsb=msbm7
\font\fivemsb=msbm5
\newfam\msbfam
\textfont\msbfam=\tenmsb
\scriptfont\msbfam=\sevenmsb
\scriptscriptfont\msbfam=\fivemsb
\edef\msbfam@{\hexnumber@\msbfam}
\def\Bbb#1{{\fam\msbfam\relax#1}}
\def\widehat#1{\setbox\z@\hbox{$\m@th#1$}%
 \ifdim\wd\z@>\tw@ em\mathaccent"0\msbfam@5B{#1}%
 \else\mathaccent"0362{#1}\fi}
\def\widetilde#1{\setbox\z@\hbox{$\m@th#1$}%
 \ifdim\wd\z@>\tw@ em\mathaccent"0\msbfam@5D{#1}%
 \else\mathaccent"0365{#1}\fi}
\font\teneufm=eufm10
\font\seveneufm=eufm7
\font\fiveeufm=eufm5
\newfam\eufmfam
\textfont\eufmfam=\teneufm
\scriptfont\eufmfam=\seveneufm
\scriptscriptfont\eufmfam=\fiveeufm
\def\frak#1{{\fam\eufmfam\relax#1}}

\csname amssym.def\endcsname

\makeatletter
\def\section{\@startsection {section}{1}{\z@}{-3.5ex plus -1ex minus
 -.2ex}{2.3ex plus .2ex}{\large\sc}}
\def\subsection{\@startsection{subsection}{2}{\z@}{-3.25ex plus -1ex minus
 -.2ex}{1.5ex plus .2ex}{\normalsize\sc}}
\makeatother

\makeatletter
\@addtoreset{equation}{section}

\makeatother

\newcommand{\nc}{\newcommand}
\newcommand{\rnc}{\renewcommand}

\nc{\be}{\begin{equation}}
\nc{\ee}{\end{equation}}
\nc{\bea}{\begin{eqnarray}}
\nc{\eea}{\end{eqnarray}}

\nc{\trac}[2]{{\textstyle\frac{#1}{#2}}}

\nc{\ex}[1]{\mbox{e}^{\,\textstyle#1}}

\nc{\CC}{\Bbb{C}}
\nc{\HH}{\Bbb{H}}
\nc{\PP}{\Bbb{P}}
\nc{\RR}{\Bbb{R}}
\nc{\ZZ}{\Bbb{Z}}
\nc{\II}{\Bbb{I}}
\nc{\EE}{\Bbb{E}}

\rnc{\a}{\alpha}
\rnc{\b}{\beta}
\rnc{\d}{\delta}
\nc{\ga}{\gamma}
\nc{\f}{\phi}
\nc{\e}{\eta}
\rnc{\c}{\chi}
\nc{\eps}{\epsilon}
\nc{\om}{\omega}
\nc{\Om}{\Omega}

\nc{\symx}{\circledS}
\newsymbol\smallsmile 1360
\newsymbol\smallfrown 1361
\nc{\ad}{\mathop{\mbox{ad}}\nolimits}
\nc{\tr}{\mathop{\mbox{tr}}\nolimits}
\nc{\Tr}{\mathop{\mbox{Tr}}\nolimits}
\nc{\Det}{\mathop{\mbox{Det}}\nolimits}
\rnc{\det}{\mathop{\mbox{det}}\nolimits}
\nc{\rk}{\mathop{\mbox{rk}}\nolimits}
\nc{\sign}{\mathop{\mbox{sign}}\nolimits}
\nc{\del}{\partial}
\nc{\diag}{\mbox{diag}}
\nc{\ra}{\rightarrow}
\nc{\Ra}{\Rightarrow}
\nc{\LRa}{\Leftrightarrow}
\nc{\lra}{\leftrightarrow}
\nc{\ot}{\otimes}
\rnc{\ss}{\subset}
\nc{\nul}{\noindent\underline}
\nc{\non}{\nonumber\\}
\nc{\mat}[4]{\left(\begin{array}{cc}#1&#2\\#3&#4\end{array}\right)}
\rnc{\lg}{\frak{g}}

\begin{document}

%%%%%%%%%%%%%%%% Definitions GP&AT %%%%%%%%%%%%%%%%%%%%%%%%%%%%%%%%%%%%%

\def\ci{\cite}
\def\ov{\over}
\def\ha{{ 1\ov 2}}
\def\four{{1 \ov 4}}
\def\td{\tilde}
\def\ff{{\rm f}}
\def\hh{{\rm h}}
\def\kk{{\rm k}}
\def\const{{\rm const}}
\def\ep{\epsilon}
\def\lc{{light-cone}}
\def\bZ{\ZZ}
\def\bR{\RR}
\def\bC{\CC}
\def\g{\gamma}
\def\foot{\footnote}
\newcommand{\rf}[1]{(\ref{#1})}
\def\la{\label}
\def \p {\phi}
\def \bi{\bibitem}
\def \s{\sigma}
\def\E{{\cal E}}
\def\H{{\cal H}}
\def\D{{\rm D}}
\def \m {\mu}
\def \aa {{\rm a}}

%%%%%%%%%%%%%%%%%%%%%%%%%%%%%%%%%%%%%%%%%%%%%%%%%%%%%%%%%%%%%%%%%%%%%%%%

%\rightline{Version of \today}

%\rightline{hep-th/0304198}
\rightline{SISSA/31/2003/EP}
\rightline{Imperial/TP/02 -3/12}
%\vskip 0.1in

\begin{center}
{\Large\sc Solvable models of strings in homogeneous plane wave backgrounds}
\end{center}

\begin{center}
{\large
M.\ Blau${}^{a}$\footnote{e-mail: {\tt mblau@ictp.trieste.it}},
M.\ O'Loughlin${}^{b}$\footnote{e-mail: {\tt loughlin@sissa.it}},
G.\ Papadopoulos${}^{c}$\footnote{e-mail: {\tt gpapas@mth.kcl.ac.uk}}
and A.A.\ Tseytlin${}^{de}$\footnote{Also at Lebedev Physics Institute,
Moscow.}\footnote{e-mail: {\tt a.tseytlin@imperial.ac.uk}}
 }
\end{center}

%\vskip 0.05 cm
\centerline{\it ${}^a$ Abdus Salam International Center for Theoretical Physics, }
\centerline{\it Strada Costiera 11, I-34014 Trieste, Italy }

%\vskip 0.05 cm
\centerline{\it ${}^b$ S.I.S.S.A. Scuola Internazionale Superiore di Studi Avanziati,}
\centerline{\it Via Beirut 4, I-34014 Trieste, Italy}

%\vskip 0.05 cm
\centerline{\it ${}^c$ Department of Mathematics,  King's College London}
 \centerline{\it London WC2R 2LS, U.K. }

%\vskip 0.05 cm
\centerline{\it ${}^d$ Theoretical Physics Group, Blackett Laboratory, }
\centerline{\it Imperial College, London SW7 2BZ, U.K.}

%\vskip 0.05  cm
\centerline{\it ${}^e$ Smith Laboratory, The Ohio State University}
\centerline{\it Columbus, OH 43210, USA}

\vskip -2.0 cm

\begin{center}
{\bf Abstract}
\end{center}
\vskip -0.1 cm

We solve closed string theory in all regular homogeneous plane-wave
backgrounds with homogeneous NS three-form field strength and a dilaton.
The parameters of the model are constant symmetric and anti-symmetric
matrices $k_{ij}$ and $f_{ij}$ associated with the metric, and a constant
anti-symmetric matrix  $h_{ij}$ associated with the NS field strength.
In the light-cone gauge the rotation parameters $f_{ij}$ have a natural
interpretation as a constant magnetic field.  This is a generalisation
of the standard Landau problem with  oscillator energies now  being
non-trivial  functions of the parameters $f_{ij}$  and $k_{ij}$.
We  develop a  general procedure for solving linear but non-diagonal
equations for string coordinates, and determine the corresponding
oscillator frequencies, the light-cone  Hamiltonian and level matching
condition.  We investigate the resulting string spectrum in detail
in the four-dimensional case and compare the results with previously
studied examples. Throughout we will find that the presence of the
rotation parameter $f_{ij}$ can lead to certain unusual and unexpected
features of the string spectrum like new massless states at non-zero
string levels, stabilisation of otherwise unstable (tachyonic) modes,
and discrete but not positive definite string oscillator spectra.

\newpage
\begin{small}
\tableofcontents
\end{small}

\newpage
\setcounter{footnote}{0}

%%%%%%%%%%%%%%%%%%%%%%%%%%%%%%%%%%%
\section{Introduction}
%%%%%%%%%%%%%%%%%%%%%%%%%%%%%%%%%%%%

A  renewal of interest  in plane wave  backgrounds
in string theory
\ci{blabla,mt,bmn,rt} prompts one to
study systematically
various  examples of solvable models of strings
moving in such ``null'' spaces, extending
early work on this subject \ci{amat,hos,veg}.
Among the   motivations (see \ci{prt} for a detailed discussion)
is a desire to learn more about
patterns of string spectra
in non-trivial curved backgrounds.

Here we will investigate  a seemingly simple but non-trivial class
of plane wave models that escaped  detailed attention
until  recently \ci{mm}. It is  based  on the
plane wave metric
\be
ds^2=2dudv + k_{ij} x^i x^j du^2 + 2 f_{ij} x^i  dx^j du+ dx^i dx^i ~ \ ,
\la{may}
\ee
supported by a null 2-form potential, $B_{ui} = h_{ij} x_j$,
and dilaton.

We shall mostly consider the model for which the matrices $k_{ij}$ and
$f_{ij}$ are constant. In that case the metric \rf{may} is regular and 
homogeneous \cite{mm}. If these matrices do not commute, $f_{ij}$ cannot  
be eliminated by a coordinate transformation while keeping the $du^2$-coefficient
$u$-independent. 
The constant matrix $f_{ij}$ can be interpreted as a ``rotation'' \ci{mm}   or
``magnetic field''  \ci{ruts} parameter.
Indeed,  the quantum mechanics of a relativistic particle propagating
in the four-dimensional plane wave with
$k_{ij} = \kk_i\d_{ij}$ and $f_{ij} = \ff \epsilon_{ij}$ in the \lc\ gauge
is  described by the Schr\"odinger equation for a two-dimensional
non-relativistic oscillator coupled  to a constant magnetic field of
strength $\ff$ with  frequencies proportional to $\kk_i$.
This is a generalisation of the standard Landau problem (for which
$\kk_i=0$). The solution of the more general problem has been given
in \ci{yonei} and references therein. For positive $\kk_i$ the \lc\ energy is
unbounded from below,  reflecting the possibility for a particle to escape
to infinity in the corresponding direction.  Switching on a non-zero $f_{ij}$
tends to stabilise the motion, trapping the trajectories near
the center of $x$-space.

We find the classical solutions of a closed string propagating in \rf{may}
in the presence of constant null NS 3-form field strength. For this we
shall develop a formalism, based on what we will call 
the frequency base ansatz, which allows for the
diagonalisation of the quantum mechanical problem.  Then we quantise
the theory in the light-cone gauge, compute the canonical commutation
relations and determine the \lc\ string spectrum. We shall
show that the frequencies depend on the parameters $k_{ij}, f_{ij}$
and $h_{ij}$ in a more  intricate way than in some previously discussed
examples    \ci{mt,bmn,rt}.  In particular, for  certain values of the
parameters there is a possibility that some  states at non-zero string
levels become  massless. We shall compute the string Hamiltonian and
investigate the level-matching condition.

Throughout we will find that the presence of the rotation parameter $f_{ij}$
leads to certain unusual and unexpected features of the string spectrum. For
instance, it may  stabilise  otherwise unstable (tachyonic) modes,
the string oscillator spectrum may be discrete but not positive
definite, and the level-matching condition for low-lying string modes
can differ from the usual form of the level-matching condition by signs.

In some special cases,  there is an  analogy with  string oscillator
spectra  found in the case of the  ``standard'' ($f_{ij}=0$) plane waves
supported by a non-constant null $B_{MN}$  background in \ci{war}.
It is possible that the metric \rf{may} supplemented  by appropriate
null background fluxes is the Penrose limit of some complicated
ten-dimensional geometries which are  ``deformations'' of the $AdS_5
\times S^5$ background.  In that case the string spectrum we shall find
below may be  relevant for comparison with its gauge-theory counterpart
along the lines of \ci{bmn} (cf.\ \ci{prez}).

The paper is organised as follows. In section two, we review some
aspects of string backgrounds with pp-wave or  plane wave metrics. Then
we turn to the special case of string backgrounds with homogeneous plane
wave metrics supported by other homogeneous fluxes like for example a
homogeneous three-form NS-NS field strength. We  discuss the solution
of the Klein-Gordon equation  for the corresponding four-dimensional
non-singular homogeneous plane wave metric, explaining its equivalence
(in the light-cone frame) to the ``2-dimensional oscillator in magnetic
field'' problem.  We investigate the properties of the spectrum of the
corresponding Hamiltonian as a function of  the parameters of the plane
wave metric.

In section three, we give the most general classical solution of a closed
string in a non-singular homogeneous plane wave background using the so
called frequency base ansatz. We illustate the procedure in the case of
four-dimensional plane waves and then go on to determine the canonical
commutation relations, the light-cone Hamiltonian and the closed string
level matching conditions in general.

In section four, we apply the above results to  study some  aspects
of  string theory in the  case of four-dimensional non-singular
homogeneous plane wave with non-vanishing homogeneous NS-NS  three-form
field strength (and/or  a dilaton).  In particular, we find the
string spectrum explicitly for various special cases, including  the
anti-Mach model \ci{mm}.  We also discuss  the explicit form of the
level matching condition and compute the zero-point energy part  of the
string Hamiltonian.

In appendix A, we summarise some facts about matrices with null vectors
and their minors which are relevant for solving string theory in a
homogeneous plane wave. In appendix B, we determine the closed string
canonical commutation relations.  In appendix C, we present a geometric
description of a degeneracy of the spectrum of some string models in
terms singularities of a hypersurface and an associated fibration. We
also compute the Berry connection of some quantum mechanical models
that arise in the context of strings in homogeneous plane waves and we
find that it vanishes.

%%%%%%%%%%%%%%%%%%%%%%%%%%%%%%%%%%%%%%%%%%%%%%%%%%%%%%%%%%%%%%%%%%%%%%%%%%%
\section{Homogeneous plane wave backgrounds
 as  a subclass of pp-wave string-theory solutions}

\subsection{pp-wave and plane wave backgrounds}
%%%%%%%%%%%%%%%%%%%%%%%%%%%%%%%%%%%%%%%%%%%%%%%%%%%%%%%

We shall begin with a review of a class of string
backgrounds with metrics admitting a covariantly constant null
Killing vector, and explain how the homogeneous plane waves discussed in
\ci{prt,mm} appear as  special cases. We shall also describe
the models that we shall focus on later in this paper.
Let us  start with the most general null Brinkmann metric
with flat transverse space
(for a review see, e.g., \ci{tcqg})\foot{If the
transverse metric is taken to be
$g_{ij}(u,x) dx^i dx^j$, then  $ K$ and $A_i$ can be eliminated by
transforming  $x^i$ and $v$; this is no longer the case
  if we set $g_{ij}
= \delta_{ij}$ (see, e.g., \ci{ts93}).}
 \be\la{met}
ds^2 = 2 dudv  + K(u,x) du^2 + 2 A_i(u,x) dx^i du\
+\  dx^i dx^i \ ,
\ee
where $i=1,\dots, d$. This is the metric of a {\em p}lane-fronted wave ($dx^i dx^i$)
with {\em p}arallel rays ($\del_v$ is covariantly constant and null), or a pp-wave
for short.

String backgrounds (conformal sigma-models) with this metric may be
constructed  by switching on a combination of ``null'' p-form field
strengths and the dilaton.
For example, one can consider string backgrounds with metric (\ref{met}),
and non-vanishing  NS-NS 2-form
gauge potential  and dilaton
\be\la{dil}
B_{iu} = B_i(u,x) \ , \ \ \ \ \ \ \ \ \   \p= \p(u) \ ,
\ee
respectively.
The one-loop sigma model conformal invariance conditions or,
equivalently, the common-sector supergravity field equations are
\bea
R_{MN}-{1\over4} H^R{}_{ML} H^L{}_{RN}+2\nabla_M\partial_N\phi&=&0
\cr
\nabla_L(e^{-2\phi} H^L{}_{MN}) &=&0 \ ,
\la{tops}
\eea
where $M,N, L=0, \dots, 9$ and $H$ is the NS-NS field strength, $H=dB$.

{}For the pp-wave ansatz above, we find
\bea
-{1\over2} \partial_i^2K+\partial_u\partial^i A_i
+{1\over 4} F_{ij} F^{ij}-
{1\over 4} H_{ij} H^{ij}
+2\partial^2_u \phi&=&0
\cr
\partial^iF_{ij}=0 \  ,
~~~~~~~~~~~~\partial^i H_{ij}&=&0~,
\eea
where $F_{ij}=\partial_i A_j-\partial_j A_i$
and $H_{ij}=\partial_i B_j-\partial_j B_i$.
The  general solution of these equations
does not define an exact  conformal sigma model,
i.e. there will be non-trivial $\a'$-corrections to the background fields.
However, all $\a'$-corrections may be absent in some special cases
(like that of  $A_i=0, B_i=0$ in \ci{amat,hos}).

One such special  choice is
\be\la{jjl}
 A_i= \pm B_i \ .
 \ee
 It  leads to a chiral null model
\ci{hts} which is an exact string solution to all orders
in $\a'$ provided the above 1-loop conditions,
\be \la{eeew}
- \ha \del_i^2 K + \del_u \del^i A_i + 2 \del^2 _u \p=0 \ ,
\ \ \ \  \ \ \ \ \    \del^i F_{ij} =0 \ , \ \
\ee
are satisfied.
Models with  $K=0, \  A_i= - \ha F_{ij} x^j, \
F_{ij} = \const, \ \p=\const$  can be interpreted
as WZW models for  non-semisimple groups \ci{NW}.

All higher-loop corrections vanish also in another special
(now ``non-chiral'') case \ci{hor,duv,ts93}
\be\la{jj}
 A_i = - \ha F_{ij}(u)  x^j\  , \ \ \ \ \ \ \
B_i =  - \ha H_{ij}(u)  x^j\  , \ee
where the string sigma model conformal invariance conditions \rf{tops}
reduce to
\be \la{coo}
- \ha \del_i^2 K +
 \four F_{ij} F_{ij} - \four  H_{ij} H_{ij}  + 2 \del^2 _u \p=0
\ . \ee
A simple special solution of the latter is
\be \la{sep}
K=  K_{ij}(u) x^i x^j  \ , \ \ \ \ \ \ \
K_{ii} =
 \four F_{ij} F_{ij} - \four H_{ij} H_{ij}  + 2 \del^2_u \p
\ .
 \ee
For this choice of $K$, the Brinkmann metric \rf{met}
is a plane wave,
 \be\la{mpw}
ds^2 = 2 dudv  + K_{ij}(u) x^i x^j  du^2 + F_{ij}(u) x^i  dx^j du\
+\  dx^i dx^i \ .
\ee
This metric is parametrised by a symmetric matrix $ K_{ij}(u)$
and antisymmetric matrix $F_{ij}(u)$.
However, there is a freedom of coordinate transformations that
in general  reduces
the amount of independent data to one {\it symmetric}  matrix
function of $u$.

Indeed, we   can make  an   {\it orthogonal}
 rotation
\be\la{pop}
x^i = M^i_j(u) y^j \ , \ \ \ \ \ \ \delta_{ij} M^i_k(u)  M^j_\ell(u) = \delta_{k\ell} \ee
that ``absorbs'' $F_{ij}$ into $K_{ij}$, giving
the plane wave metric in Brinkmann coordinates,
 \be\la{opw}
ds^2 = 2 dudv  + \td  K_{ij}(u) y^i y^j  du^2 +
dy^i dy^i \ ,
\ee
where (a prime denotes a $u$-derivative)
\be
\la{redd}
M'^i_j =   \ha  F_{ik} M^k_j  \ ,\ \ \ \ \ \ \ \
\td  K_{ij} =  M^k_i M^l_j  (K_{kl} - { 1 \ov 4} F_{kn} F_{ln})
\;\;.
\ee
Note that  $ \td K_{ii} = K_{ii} -  { 1 \ov 4} F_{ik} F_{ik}$
which is in agreement with the expression for the Ricci tensor component
$R_{uu}$ in \rf{coo},\rf{sep}.
If $K_{ij}$ were constant, $\td K_{ij}$
would  be $u$-dependent even for a constant $F_{ij}$.
The  rotation \rf{pop} also affects the expression for
$B_i$ in \rf{jj}.

We can also  transform  the metric \rf{mpw} or \rf{opw}
to Rosen coordinates  $(u,V, X^a)$
\be
\la{ros}
x^i = L^i_a(u) X^a \ ,\ \ \ \ \ \ \ \ \
v= V  + \ha  s_{ab} (u) X^a X^b    \ ,  \ee
getting
\be \la{roo}
ds^2 = 2 du dV +  g_{ab}(u) dX^a dX^b   \ , \ \ \ \ \ \ \ \
 g_{ab} =  L^i_a(u)  L^i_b(u)  \ . \ee
Starting with \rf{opw}, we  find \rf{roo}  provided
\be   s'_{ab} + \td  K_{ij}  L^i_a  L^j_b +  L'^i_a  L'^i_b =0
\ , \ \ \ \ \
 s_{ab}  +  L^i_{(a} L'^i_{b)} =0 \ ,  \ \ {\rm i.e. } \ \ \
  L^{i}_a{}^{''} = \td  K_{ij}   L^j_a  \ . \ee
For other aspects of the relation between Brinkmann and
Rosen coordinates for plane waves, see,  e.g., \ 
the discussion in \cite{mm}.

%%%%%%%%%%%%%%%%%%%%%%%%%%%%%%%%%%%%%%%%%%%%%%%%%%%%%%%%%%%%%
\subsection{Homogeneous plane wave backgrounds}
%%%%%%%%%%%%%%%%%%%%%%%%%%%%%%%%%%%%%%%%%%%%%%%%%%%%%%%%%%%%

Amongst the various conformal
plane wave backgrounds \rf{mpw} with $F_{ij}\not=0$,
there are two special classes associated with the homogeneous
plane wave metrics of \cite{mm}, namely
\be\la{one}
I: \ \ \   K_{ij}(u)=k_{ij} = \const \ , \ \ \ \ \ \
F_{ij}(u) = 2f_{ij}=\const  \ ,\ \ \ \ \ \
H_{ij}(u) = 2h_{ij} =\const \ ,
\ee
\be
\la{two}
II:  \ \ \ \ \ \ \ \ \ \ \ \ \ \ \ \ \ \
K_{ij}(u) ={1 \ov u^2} k_{ij}  \ , \ \ \ \ \ \ \ \ \ \ \
F_{ij}(u) = {2 \ov u} f_{ij}  \ , \ \ \ \ \
\ \ \ \ \ \  H_{ij}(u) = {2 \ov u} h_{ij}  \ .
\ee
The homogeneity of the plane wave metrics is not always respected
by the full string background. The
NS-NS fluxes are not homogeneous unless the matrices $f$ and $h$ commute,
$[f,h]=0$.
The presence of a non-constant dilaton also breaks homogeneity. 
The metric of the class I background is smooth while that of the
class II background is singular, i.e.\ these
two classes can be distinguished as non-singular and singular homogeneous
plane wave backgrounds,  respectively.

According to \rf{redd}, the   matrix $f\not=0$  can
be associated with a coordinate rotation.
For that reason,  we shall sometimes call $f$ a ``rotation'' matrix;
its other natural interpretation  (in the \lc\ gauge)
is as an effective magnetic field parameter \ci{ruts}.

We can rewrite the metrics for  the two
backgrounds (\ref{one}) and (\ref{two}) in Brinkmann coordinates
\rf{opw} as
\be
ds^2=2dudv+ (e^{-fu} \td k e^{fu})_{ij} y^i y^j du^2+  dy^i dy^i  \  ,
\la{metone}
\ee
\be
ds^2=2dudv+ u^{-2}(e^{-f\ln u } \td k e^{f\ln u})_{ij} y^i y^j du^2+ dy^i dy^i
\ , \la{mettwo}
\ee
where $ \td k_{ij}  = k_{ij} -  f_{ik} f_{jk}$, and
$y_i=(e^{ -f u})_{ij} x_j , $  $y_i=(e^{-f\ln u})_{ij} x_j $, respectively.
Among all the metrics of the form (\ref{opw}), these backgrounds are singled
out by the existence of an isometry with a non-trivial component in the 
$u$-direction. In the ``stationary'' coordinates (\ref{mpw}), the corresponding
Killing vectors are simply $\del_u$ and $u\del_u-v\del_v$ respectively.  

It is clear that if $f$ commutes with $k$, $[f,k]=0$,  the $u$-dependent
rotation factors cancel out, i.e. we can simply absorb the dependence
on the matrix $f$ into the shift $k_{ij} \to \td k_{ij} $.  This
transformation leaves the two-form NS gauge potential $B$ invariant.
The  models of class I with $f=0$ have been widely investigated in the
literature (see \ci{prt} for references and \ci{sadri} for a more recent
investigation of some particular models with $f=0$).

For generic matrices $k,f,h$, there is still some redundancy in the
parametrisation of the backgrounds. In both cases  I and II, it is always
possible to diagonalise $k$ with an $O(d)$ orthogonal transformation of
the coordinates $x$. In addition, under the scaling transformation $u
\to \ell u, \ v \to \ell^{-1} v$, the parameters of the models of class
I transform as $k\to \ell^2 k$, $f\to \ell f$ and $h\to \ell h$. Thus
the eigenvalues of $k$ should be specified up to a positive constant.
The remaining  $Z_2$ symmetry ($\ell=\pm 1$)  allows  one to determine
$f,h$ up to a sign.

The scaling transformation $u \to \ell u, \ v \to \ell^{-1} v$ is
a symmetry of the models of class II. In fact, the presence of this
additional isometry is responsible for the homogeneity property of the
corresponding  metrics.  These models are parametrised by the eigenvalues
of $k$ and the two skew-symmetric matrices $f,h$.  Models with $f=0$
have been investigated  in \ci{prt} and models with $f\not=0$ have been
considered in \ci{mm}.

Our aim here is to investigate string theory on the
backgrounds (\ref{one}) and (\ref{two}) with  $f_{ij}\not=0$ and $[k,f]\not=0$.
In both cases, the sigma model conformal invariance condition (\ref{coo})
can be solved in terms of the dilaton to give
\be \la{dio}
I:\ \ \  \  \p=\p_0+c\,u- \ha \m u^2
\ee
\be\la{dit}
II:\ \ \  \  \p=\p_0+c\,u +  \m \ln u
\ee
where
\be\la{dim}
\m= - \ha (k^i{}_i- f_{ij} f^{ij}+h_{ij} h^{ij})~
\ee
and $\p_0$ and $c$ are  arbitrary  integration constants.

An important condition on our models is
the requirement that the  string coupling $e^\phi$
be small for all values of $u$.  In  the case I, we find that
this leads to the requirement
\be\la{ooi}
\m\geq 0~.
\ee
For the models of class II with $\m\not=0$, small string coupling requires
that $c<0$ and $\m >  0$ for $u\geq 0$. A more detailed discussion of
this issue can be found in \cite{prt} for a similar model  with $f=0$.

The condition  \rf{ooi} is satisfied automatically
if $f_{ij}\not=0$ but $k_{ij}=0$ and $h_{ij}=0$: then
$\m= \ha f_{ij} f^{ij}$. Since in this case $[k,f]=0$,
we can perform a coordinate transformation to construct a model with
$\td k=f^2$ and $f=0$.\foot{Note that only a subclass of models
with $k\not=0$ and $f=0$
can be constructed in this way because not every symmetric matrix
is the square of a skew-symmetric one.
Conversely, if $k=\ff^2$ for some skew-symmetric
matrix $\ff$ and $f=0$, then after a coordinate transformation
such a model is  related to one with $k=0$ and $f=\ff$.}

%%%%%%%%%%%%%%%%%%%%%%%%%%%%%%%%%%%%%%%%%%%%%%%%%%%%%%%%%%%%%
\subsection{Four-dimensional homogeneous plane waves
with rotation}
%%%%%%%%%%%%%%%%%%%%%%%%%%%%%%%%%%%%%%%%%%%%%%%%%%%%%%%%%%%%%%%%%%%%%%%%%%%

The simplest examples of homogeneous plane wave backgrounds
with rotation for which $k\not=0$, $f\not=0$ and $[k,f]\not=0$
are  four-dimensional ones, where $ k$ and $f$ are
$2 \times 2$ matrices.

The most general 4-d model of that type \rf{jj}, (\ref{mpw})
with parameter functions in \rf{one},\rf{two}
has
\be\la{od}
k_{ij} =   {\rm diag}({\kk_1, \kk_2 })  \ , \ \ \ \
f_{ij} =  \ff \ep_{ij}  \ , \ \ \ h_{ij}= \hh \ep_{ij}  \ .
\ee
If $\kk_1=\kk_2$, then $[k,f]=0$ and the rotation matrix
$f$  can be set to zero
by  a coordinate transformation. 
For that reason we shall
focus on  the non-trivial case where
$$\kk_1\not= \kk_2 \ .   $$
For models with constant dilaton, we have to restrict the parameters as
\be\la{coon}
\kk_1 + \kk_2 =  2\ff^2 -  2\hh^2 \ .
\ee
We may instead set the 2-form NS-NS gauge potential  to zero
$(h_{ij}=0)$ and introduce a dilaton, i.e. consider the following models
 \be\la{mood}
k_{ij} = {\rm diag}({\kk_1, \kk_2})  \ , \ \ \ \
f_{ij} = \ff \ep_{ij}  \ , \ \ \ h_{ij}=0 \ ,
\ee
with the dilaton given by \rf{dio}  or \rf{dit} with
\be\la{kop}
  \m=-{1\over2}(\kk_1 + \kk_2) + \ff^2 \ .
 \ee

%%%%%%%%%%%%%%%%%%%%%%%%%%%%%%%%%%%%%%%%%%%%%%%%
\subsection{Models with constant dilaton}
%%%%%%%%%%%%%%%%%%%%%%%%%%%%%%%%%%%%%%%%%%%%%%%%%%%%

A subclass of models for which the dilaton $\phi$ is constant
is found for  $\m=0$ and  $c=0$.
{}For $h=0$, the condition of
$\m=0$ implies (in both cases I and II) that
\be \la{kkk}
k^i{}_i =  f_{ij} f^{ij} \ .
\label{onon}
\ee
A  solution of (\ref{onon}) with constant parameters is
\be
\la{mah}
k_{ij} =    {\rm diag}({2\ff^2, 0 })  \ , \ \ \ \
f_{ij} =  \ff \ep_{ij} \ . \ee
The corresponding metric  is the anti-Mach metric (see \ci{mm} and section 4.2 below).
Here $\kk= 2 \ff^2$ can be set equal to 1 by a rescaling of $u$ and $v$.

A ten-dimensional  solution can be constructed by setting
$$
f=\ff (\epsilon\oplus \epsilon\oplus \epsilon \oplus \epsilon)
\ , \ \ \ \ \ \ \ \
\td k = k + f^2 ={\bf 1}_4\oplus (-{\bf 1}_4)\ .
$$
This background
is characterised by an additional $U(2)\times U(2)\times \bZ_2$ symmetry
acting on the coordinates $y$ in (\ref{metone}).
To see this we can  write the transverse space as $\bR^8=\bC^2\oplus\bC^2$
using the obvious complex structure defined by $f$.
Each $U(2)$ acts as a holomorphic
isometry on each $\bC^2$. The non-trivial element of $\bZ_2$ exchanges the
two $\bC^2$ subspaces.

Let us also mention that it is straightforward  to construct
similar plane wave models containing in addition  null
R-R  background  field strengths.
The corresponding string models  can be solved using the GS formalism
in the \lc\ gauge, as was done in  \ci{mt,bmn,rt}.
The effect of the R-R parameters on the metric (for constant dilaton) can
be taken into account by replacing $H_{ij}^2$ in \rf{sep}
(and thus $\hh^2$ in \rf{coon}) by the sum of the squares of all null field strengths.

%%%%%%%%%%%%%%%%%%%%%%%%%%%%%%%%%%%%%%%%%%%%%%%%
\subsection{World-sheet fermionic couplings in the \lc\ gauge}
%%%%%%%%%%%%%%%%%%%%%%%%%%%%%%%%%%%%%%%%%%%%%%%%%%%%

To generalise the discussion to the superstring case, we note
that the world-sheet (NSR)  fermions decouple  from the
function  $K_{ij}$ of the  plane-wave  metric \rf{mpw} in the \lc\ gauge
$u= p_v \tau, \ \psi^u=0$\  ($2\a'=1$).
In the presence of the NS-NS 2-form field, the coupling to the
functions $F_{ij}$ and $H_{ij}$ is (for details see, e.g., \ci{rt})
\be
\la{fer}
 L_F= \psi^i_L [ \d_{ij} \del_- -  \four   p_v (F_{ij} -  H_{ij} )]
\psi^j_L + \psi^i_R [ \d_{ij} \del_+ -  \four  p_v  ( F_{ij} + H_{ij} )]
\psi^j_R \ . \ee
The   dependence on $F_{ij}(u)$   can be eliminated
by a local  Lorentz  transformation (rotation of fermions)
as in \rf{redd}. Indeed,
we can apply the coordinate transformation \rf{pop}  to eliminate $F_{ij}$
by transforming $K_{ij}$ into $\td K_{ij}$ \rf{redd} which will
not couple to the fermions.
Thus an equivalent fermionic Lagrangian is
\be \la{ferm}
 L_F= \psi^i_L ( \d_{ij} \del_-  +   \four   p_v   H_{ij} )
\psi^j_L + \psi^i_R ( \d_{ij} \del_+ -  \four  p_v  H_{ij} )
\psi^j_R \ . \ee
To conclude,   the    fermions do not couple to the  parameters of the
metric \rf{mpw} in the \lc\ gauge. This is consistent with the fact that
the fermions do not contribute to (1-loop)  conformal invariance conditions in this
case.

We can also apply another local Lorentz rotation
to get another equivalent action where the left-moving  fermions are free
\be \la{frm}
 L_F= \psi^i_L   \del_- \psi^i_L + \psi^i_R ( \d_{ij} \del_+ -  \ha  p_v  H_{ij} ) \psi^j_R \ . \ee
Similar \lc\ gauge actions are found in the Green-Schwarz approach.
For example, the analog of  \rf{ferm} is
\be \la{gs}
 L_F= i S_L (
\del_- + {1 \ov 16} p_v H_{ij} \g^{ij} ) S_L + i S_R ( \del_+
- {1 \ov 16} p_v  H_{ij} \g^{ij} ) S_R \ .
 \ee

%%%%%%%%%%%%%%%%%%%%%%%%%%%%%%%%%%%%%%%%%%%%%%%%%
\subsection{Solution of the four-dimensional Klein-Gordon equation}
%%%%%%%%%%%%%%%%%%%%%%%%%%%%%%%%%%%%%%%%%%%%%%%%%%%%%%

In section 3 we will explain a general procedure for quantising particles
and strings in regular homogeneous plane wave backgrounds. The case of the
relativistic particle in four dimensions (two transverse dimensions) can
also be solved directly, and for comparison purposes we present this
solution here.

The quantisation of a relativistic particle  propagating in the
homogeneous non-singular plane wave background
\be
ds^2=2dudv+ k_{ij} x^i x^j du^2 + 2 f_{ij}x^i  dx^j du+ dx^i dx^i ~ \ ,
\la{mmm}
\ee
leads to the  Klein-Gordon equation.
In general,  the  massive Klein-Gordon equation
$( \nabla^2 - M^2) \Psi=0$  corresponding to  the metric
\rf{met} has the form
\be
\la{kg}
\big[2\partial_u \partial_v-K \partial_v^2+(\partial_i-A_i\partial_v)^2  -
M^2\big]\Psi(u,v,x) =0 \;\;.
\ee
When $K$ and $A_i$ do not depend on $u$
(as is the case for the  homogeneous plane wave
background) it is natural to  perform a Fourier transform in
the $u,v$ coordinates,
$$
\Psi (u,v,x) = \int dp_v dp_u e^{ i p_u u + i p_v v}  \psi (p_u, p_v;x) \;\; .
$$
Then  the equation \rf{kg} becomes the Schr\"odinger-type
 equation for
 $\psi(p_u, p_v,x)$
\be
H \psi=\E\psi\;\;,\;\;\;\;\;\;
\E\equiv -\ha M^2 - p_u p_v\;\;,
\ee
where
\be
\la{hmmm}
H= -\ha[(\partial_i-i p_v A_i)^2+ p_v^2 K(x)]
\ee
may be interpreted as a non-relativistic Hamiltonian of a particle with charge
$p_v$ coupled to a magnetic field  $A_i$ and moving in
a potential  $V=- p_v^2 K(x)$. The light-cone
Hamiltonian is
\be
H^{(0)}=-p_u = \frac{M^2}{2p_v} + p_v^{-1}H\;\;.
\ee
In the case of the homogeneous plane wave in four dimensions \rf{mood},
we have
$$
A_i=-\ff \epsilon_{ij} x^j\;\;,\;\;\;\;\;\;
K=\kk_1 x_1^2+\kk_2 x_2^2 \;\; \ .
$$
Setting $x=x^1$ and $y=x^2$, we get the Hamiltonian operator
\be
\la{hah}
H=-\ha [(\partial_x+ip_v \ff  y)^2+(\partial_y-ip_v \ff  x)^2
+p_v^2(\kk_1 x^2 +\kk_2 y^2)]\;\;.
\ee
This  is recognised as a Hamiltonian describing the dynamics  of a
non-relativistic charged  2-dimensional oscillator (with masses $m_i=1$,
frequencies $ \omega_i^2 = - p_v^2 \kk_i$ and charge $p_v$) coupled to a
constant magnetic field of strength $\ff$. The solution of this problem
is well known (see, e.g.,  \ci{yonei} and references therein). Such a
Hamiltonian  can be related to a  standard  decoupled two-dimensional
harmonic oscillator Hamiltonian by a unitary transformation,
\be
H= U(\theta_1, \theta_2) (H_1+H_2) U^+(\theta_1, \theta_2) \ ,
\ee
where
\be
\la{uni}
U(\theta_1, \theta_2)=e^{-i \theta_1   x y}
e^{-i \theta_2
{p_x p_y}} \ , \  \ \
\ \ \ \ \  p_x = -i \del_x\ ,  \     \ \ p_y = - i \del_y \ ,   \ee
and
\be
\la{hmm}
H_i= {1\over 2 m_i} p_i^2+ {m_i\over2} \td{\omega}^2_i x^2_i~.
\ee
The parameters  $\theta_1, \theta_2$ are
\bea\la{tat}
\theta_1= -p_v { (\kk_1-\kk_2)+\D\over 4 \ff}
\ , \ \ \ \ \ \ \ \ \ \
\theta_2= - {2 \ff\over p_v \D} \ ,
\eea
where
\be
\la{daa}
\D^2=(\kk_1-\kk_2)^2 -   8 \ff^2  (\kk_1+\kk_2)+16 \ff^4
= (4  \ff^2 - \kk_1- \kk_2 )^2 - 4 \kk_1\kk_2\ .
\ee
The frequencies and masses  of the free harmonic oscillators are
$\td{\omega}_i= p_v \omega_i$, with
\be
\omega^2_{1,2}= \ha  (4\ff^2-\kk_1-\kk_2  \mp \D)\;\;,
\la{freq}
\ee
and
\be
\la{maas}
m_{1,2}={2\D\over \D+\kk_1-\kk_2\mp 4\ff^2}\;\;.
\ee
We are assuming that $\D$ is real and positive and  that $ \kk_1 > \kk_2$;
otherwise the labels of the two frequencies  should  be interchanged
(for $\ff=0$ we have $\omega^2_i = - \kk_i$).  The above expressions
apply when $\theta_i$ and $m_i$ are finite. When the $m_i$ are finite,
they can be eliminated by a canonical transformation generated by
$x_i \ra x_i |m_i|^{1/2}$, so the spectrum only depends on the sign of $m_i$.
The situation when $m_i=0$ or $\infty$ requires special consideration.

We will rederive these frequencies in a different way in the next
section. We will also see that in the {\em frequency basis} adopted there
(see section 3.2) the particle or string Hamiltonian is automatically a
sum of decoupled harmonic oscillators (section 3.6) without the need to
perform explicitly the counterpart of the above unitary transformation.
Various other features of this spectrum and its string mode counterpart
will be discussed in section 4.

%%%%%%%%%%%%%%%%%%%%%%%%%%%%%%%%%%%%%%%%%%%%%%%%%%%%%%%%%%%%%%
\section{Strings in non-singular  homogeneous plane-wave backgrounds}
%%%%%%%%%%%%%%%%%%%%%%%%%%%%%%%%%%%%%%%%%

\subsection{The classical string mode equations}
%%%%%%%%%%%%%%%%%%%%%%%%%%%%%%%%%%%%%%%

Our aim   is to solve the classical equations of string theory in
non-singular (case I in \rf{one}) $2+d$ dimensional
homogeneous plane-wave backgrounds  with metric (\ref{mmm}),
\be\la{mou}
ds^2=2dudv+ k_{ij} x^i x^j du^2 + 2 f_{ij}x^i  dx^j du+ dx^i dx^i ~ \ ,
\ee
where $k_{ij}$ and $f_{ij}$ are constant,
and  $B$-field as in \rf{dil}, \rf{jj} and \rf{one},
i.e.
\be
\la{bee}
 B_{iu} = - h_{ij} x^j \;\;.
\ee
The dilaton background does not enter the classical string equations of
motion but modifies the quantum stress tensor, so  the discussion below
will apply to cases of both constant and non-constant dilaton.

We shall denote the string embedding coordinates  associated
with the spacetime coordinates $(u,v,x^i)$
as $X^M=(U,V,X^i)$. Choosing the orthogonal gauge for the world-sheet metric,
the standard sigma model Lagrangian
\be
\label{Lag}
L=\frac{1}{4\pi\alpha'}(g+B)_{MN} \partial_+ X^M \partial_- X^N
\ee
(we will set $\a'= \ha$ in the following)
leads to the following  equations for $U$ and $X^i$:
\be
\partial_+\partial_-U=0\ ,
\la{ueqn}
\ee
\be
\la{xxx}
-\partial_+\partial_-X_i+ ( f+h)_{ij}\partial_-U \partial_+X^j
+ ( f-h)_{ij} \partial_+U \partial_-X^j+ k_{ij}X^j \partial_+U \partial_-U=0
\ , \ee
where $\sigma^\pm=\tau\pm \sigma$,  $ \del_\pm =   \del_\tau \pm \del_\sigma$.
The equation for $U$  is solved
by $U=U(\sigma^+)+ U(\sigma^-)$; after gauge fixing the world-sheet
conformal transformations by the \lc\ gauge this becomes
\be
\la{uuu} U=p_+ \sigma^+    +p_-\sigma^-  \;\;.
\ee
In the case of non-compact $U$-direction
(the case we shall consider below),
the condition of periodicity of $U$
in $\s$ implies that $p_+=p_-$, i.e. $U= 2 p_+ \tau$.
In most of what follows, we shall set
\be
\la{pii}
p_+ =p_-= \ha p_v  \ ,
\ee
except in some  equations below which for generality we write down
as if $p_+ $ and $p_-$ were independent to allow for compact $U$ in which case
$p_+$ and $p_-$ are quantised.

Substituting \rf{uuu} into the equation \rf{xxx} for $X^i $, we find
\be
-\partial_+\partial_-X_i+2 p_- (f+h)_{ij} \partial_+X^j
+ 2p_+ (f-h)_{ij} \partial_-X^j+4 p_+ p_- k_{ij} X^j=0
\label{xeqn}
\ee
while $V$ can be expressed in terms of $X^i$ using the equations
associated with the variation of the 2-d metric.

To solve (\ref{xeqn}), we set
\be
X^i = \sum_{n=-\infty}^{+\infty} X_n^i(\tau)~
\ex{2in\sigma} \ ,   \ \ \ \ \ \ \ \ \  X^i_n = (X^i_{-n})^*  \ ,
\ee
where $0 < \s  \leq \pi$ and  $\a'=\ha $, which leads to
$$
-\ddot  X_n^i +  2 p_- (h_{ij}+f_{ij}) \dot  X_n^j
+2 p_+ (-h_{ij}+f_{ij})  \dot  X_n^j  +
 4 (p_+ p_- k_{ij}-n^2\delta_{ij}) X_n^j  $$ \be
+ \ 4n i p_- (h_{ij}+f_{ij}) X_n^j-4n i p_+ (-h_{ij}+f_{ij}) X_n^j=0\ .
\la{timeqn}
\ee
Setting $p_+=p_-=\ha p_v$, this equation becomes
\be
-\ddot  X_n^i +  2 p_v f_{ij} \dot  X_n^j + (p_v^2 k_{ij}-4 n^2\delta_{ij}) X_n^j
+ 4 in  p_v h_{ij}X_n^j=0\;\;.
\label{ddx}
\ee
For $n=0$ this equation is that of a relativistic particle
and, as expected, the dependence on $h_{ij}$ drops out.

For the metrics of class II in \rf{two}, the analogue of this equation is
\be
-\ddot  X_n^i +   \frac{2 p_v}{\tau}f_{ij} \dot  X_n^j +
(\frac{p_v^2 k_{ij} + p_v f_{ij}}{\tau^2}-4 n^2\delta_{ij}) X_n^j
+ \frac{4 in  p_v}{\tau} h_{ij}X_n^j=0\;\;.
\label{ddxII}
\ee
For $n=0$ and changing $\tau$ to $t=$log $ \tau$ \  this is identical to
the equation of the relativistic particle in the metrics of class I.
For general ($n\neq 0$) string modes, the class II metrics have two
easily analysable limits:
\begin{itemize}
\item
When $\tau \rightarrow 0$ for any $ n$, the same
change of variables to logarithmic time $t$ reduces these equations to
the same form as the particle case of class I,  containing, however,  an
interesting correction in the constant term proportional to $f_{ij}$.
This case can be studied using the same
methods that we develop below, and then this limit becomes the basis
for an expansion of the solution around $\tau = 0$.
\item
When $\tau \rightarrow \infty$ all
$\tau $ dependent terms go to zero and we find just a string in flat space.
\end{itemize}

In  the remainder of this paper we will concentrate exclusively on the
non-singular homogeneous plane wave solutions of class I.

As explained  above, without loss of generality we can assume that $k_{ij}$ is diagonal,
$k_{ij}=\kk_i \d_{ij}$.
Also, from now on, to simplify the notation, we will set
$$p_v=1 \ . $$
The dependence
of any equation on $p_v$ can be reinstated by scaling $k_{ij}\ra p_v^2 k_{ij},
f_{ij}\ra p_v f_{ij}, h_{ij}\ra p_v h_{ij}$.

%%%%%%%%%%%%%%%%%%%%%%%%%%%%%%%%%%%%%%%%%%%%%%%%%%
\subsection{The Frequency Base ansatz}

The general method to solve  a system of $d$ coupled second order
equations  like \rf{ddx}
is to rewrite it as a set of $2d$ first-order equations. This
first-order $(2d \times 2d)$ matrix differential equation with constant
coefficients is then solved by exponentiating this matrix. In practice,
however, this is rather unwieldy.

Fortunately,  for generic values of the parameters appearing in (\ref{ddx}) one
can use a much simpler procedure. Namely, to solve these equations,
one makes the ansatz
\be
X^i_n = \sum_{J=1}^{2d} \xi^{(n)}_{J} a^{(n)}_{iJ}\ex{i\omega^{(n)}_J \tau}\;\;,
\label{xfb}
\ee
with the frequencies $\omega^{(n)}_J$ and their eigen-directions $a^{(n)}_{iJ}$
to be determined.\footnote{At this stage the $\xi$'s are redundant as they
could be absorbed into the $a_{iJ}$. They
have been introduced here for later convenience, as they will be promoted
to operators upon quantisation. Their commutation relations will be
determined from the canonical commutation relations of the $X^i$ and their
 conjugate momenta.}

Plugging this into the differential equation (\ref{ddx}), one
obtains the matrix equation
 \be M_{ik}(\omega^{(n)}_J,n)\ a^{(n)}_{kJ}\
=0\;\;,
\ee
where \be M_{ik}(\omega,n) = (\omega^2 + \kk_{i}-4
n^2)\delta_{ik} + 2 i \omega f_{ik} + 4 in h_{ik}\;\;.
\label{Mik}
\ee
A necessary condition for finding a solution to this matrix
equation is that
\be
\det M(\omega,n)= 0\;\;.
\label{detcon}
 \ee
This
equation has $2d$ roots $\omega = \omega^{(n)}_J$,
$J=1,\ldots,2d$, which are the frequencies entering the ansatz
(\ref{xfb}). If all the roots are found to be distinct, or if
equal roots are associated with linearly independent null
eigenvectors, then the frequency base ansatz is justified because
the expansion (\ref{xfb}) then involves all the $2d$ linearly
independent solutions of the equation (\ref{ddx}). The degenerate
case when there are multiple roots associated with the same
eigenvector requires separate considerations. Unless stated
otherwise, in the following we will always assume that all the
roots are  distinct.

We will see below that in the generic case this frequency base ansatz
is very efficient, in particular, because in this basis the canonical
commutation relations and the string Hamiltonian will turn out to be
automatically diagonal (a sum of decoupled harmonic oscillators).

We also want to mention that,  in addition to the general $(2d\times 2d)$ matrix method
and the frequency base ansatz,  there is also a third, in some sense,
 intermediate,
possibility:  to make the ansatz
\be
X_n(\tau) = \ex{\tau\Lambda_n}X_n(0)
\ee
with $\Lambda_n$ being  some $(d\times d)$ matrix.
This leads to a quadratic matrix equation
for $\Lambda_n$, and works well in certain simple cases. However, this method
is neither as general as the first one, nor as efficient as the second, and we
will not make use of it in the following.

%%%%%%%%%%%%%%%%%%%%%%%%%%%%%%%%%%%%%%%%%%%%%%%%%%%%%%%%%%%%%%
\subsection{Warm-up: four-dimensional plane waves}

Before explaining in general how to obtain the solutions for the string
modes, we illustrate the procedure in the case of four-dimensional homogeneous
plane waves. Various other properties of these four-dimensional plane waves
will be discussed in detail in section 4.

For two transverse dimensions, the matrix $M$
takes the form
\be\la{mam}
M = \mat{\omega^2 + \kk_1 - 4 n^2}{2i\ff\omega +
4 i n \hh}{-2i\ff\omega -4in\hh}{\omega^2 + \kk_2 - 4n^2}
\ee
where we set $f_{ij}=\ff \epsilon_{ij}$ , $h_{ij}=\hh \epsilon_{ij}$
in \rf{Mik}.
Thus the equation for the roots (frequencies) is
\be
\omega^4 + (\kk_1 + \kk_2 - 4\ff^2 - 8 n^2) \omega^2  - 16n \ff\hh \omega
+ (\kk_1-4n^2)(\kk_2-4n^2)
- 16 n^2 \hh^2 = 0\;\;.
\label{4droots}
\ee
We see that this equation is invariant under $(\omega,n)\ra(-\omega,-n)$.
In particular, the roots $\omega_{J}^{(0)}$ for $n=0$ come in pairs $\pm \omega_j$,
and for each frequency $\omega_{J}^{(n)}$ of the $n$'th string mode
there is a corresponding frequency $\omega_{J}^{(-n)}=-\omega_{J}^{(n)}$ of the
$(-n)$'th mode.

For $n=0$, this becomes a quadratic equation for $\omega^2$,
\be
\omega^4 + (\kk_1 + \kk_2 - 4\ff^2) \omega^2  + \kk_1\kk_2 = 0\;\;,
\label{ofour}
\ee
with solutions\footnote{The frequencies
have been labelled in such a way that for $\ff=0$ one finds the standard result
$\omega_i^2 = - \kk_i$.}
\be
\omega_{1,2}^2 = \frac{1}{2}(4\ff^2-\kk_1-\kk_2) \mp
\frac{1}{2}\sqrt{(4\ff^2-\kk_1-\kk_2)^2 - 4 \kk_1\kk_2}\;\;.
\label{fbf}
\ee
Note that these  are precisely the frequencies (\ref{freq}) obtained in
section 2.6 from the explicit diagonalisation of the particle Hamiltonian.
This will also be  verified later  when we compute the
Hamiltonian operator using the  classical solutions of
the system.

We obtain four distinct frequencies unless either
the discriminant $\D$ (given by the same expression as in \rf{daa})
\be
\D^2=(4\ff^2-\kk_1-\kk_2)^2 - 4 \kk_1\kk_2 \;\;,
\ee
vanishes, or one has $\kk_1\kk_2 = 0$.

Similarly, for the string modes ($n \neq 0$),  and when either $\hh=0$ or $\ff=0$,
one obtains a quadratic equation for $\omega_J^2$. In the former
case the solutions can be obtained
from the solutions for $n=0$ by shifting $\kk_i \ra \kk_i - 4 n^2$, so
that the frequencies are
\be\la{fern}
(\omega^{(n)}_{J})^2 = \frac{1}{2}(4\ff^2
 -\kk_1-\kk_2+ 8 n^2) \mp \frac{1}{2}\sqrt{(4\ff^2  -\kk_1-\kk_2+ 8 n^2)^2 - 4
(\kk_1-4n^2)(\kk_2-4n^2)}\;\;.
\ee
In the latter case, we find that
\be\la{hfo}
(\omega^{(n)}_{J})^2=\frac{1}{2}(
 -\kk_1-\kk_2+ 8 n^2) \mp \frac{1}{2}\sqrt{(\kk_1-\kk_2)^2 +64 n^2 h^2}\;\;.
 \ee
The squares of the frequencies are real for physical values of
the parameters.

In the general case with $n\ \hh\  \ff \neq 0$, the equation for the
frequencies is a quartic (with vanishing cubic term), whose roots can
of course be given in a closed (albeit complicated and unenlightning) form.

Having found the frequencies $\omega_J$, we now need to determine  the
corresponding null eigenvectors $a_{iJ}$.
These can readily be constructed explicitly on a case-by-case basis,
but there is also a general method  which we outline now (and which,
suitably interpreted, immediately generalises to the higher-dimensional
case). Namely, we observe that for any $(2 \times 2)$ matrix $M$ with
vanishing determinant,
\be
M_{11}M_{22}-M_{21}M_{12} = 0\;\;,
\ee
a possible choice for the corresponding null eigenvector $v_k$ with
$M_{ik}v_k=0$ is
\be
(v_1,v_2)=(-M_{22},M_{21})\;\;.
\ee
Indeed,  one then has
\bea
M_{1k}v_{k} &=& - \det M = 0\non
M_{2k}v_{k} &=& -M_{21}M_{22} + M_{22} M_{21}= 0\;\;.
\eea
Therefore,  we can  choose the null eigenvectors $a_{kJ}$
to be, e.g.,
\be
(a_{1J},a_{2J})=(-M_{22}(\omega_J),M_{21}(\omega_J))\;\;.
\label{4dai}
\ee
This is an acceptable choice of null eigenvector unless
$M_{2i}(\omega_J)\equiv 0$, in which case one can use
$M_{1i}(\omega_J)$ instead. If $M(\omega_J)$ is identically
zero, obviously the construction of null eigenvectors in terms of matrix elements
of $M$ fails. But the construction of (two linearly independent)
null eigendirections is trivial in this case.

We have now completely determined the frequency expansion (\ref{xfb})
of the string modes, up to an overall scale encoded in the $\xi$'s
parametrising the classical solutions.

%%%%%%%%%%%%%%%%%%%%%%%%%%%%%%%%%%%%%%
\subsection{The general classical solution}

We now discuss the construction of the solutions (\ref{xfb}) in
the general case.
First of all, an important property of $M$ in \rf{Mik}
is that
\be
M^T (\omega,n)= M(-\omega,-n)\;\;.
\ee
Thus $M(\omega,n)$ and $ M(-\omega,-n)$
have the same determinant and hence the same roots.
This means that for $n=0$ the frequencies come in $\pm$ pairs,
\be
n=0:\;\;\;\;\{\omega_J\} = \{\pm \omega_j, \ \ j=1,\ldots,d\}\;\;,
\ee
as we had already seen in the four-dimensional case.
It is then convenient to rewrite the expansion of the zero-mode as
\be
X^i_0 = \sum_{j=1}^{d} \ [\xi^{+}_{j} a^{+}_{ij}\ex{i\omega_j \tau}
+ \xi^{-}_{j} a^{-}_{ij}\ex{-i\omega_j \tau} ]\;\;.
\ee
For $n\neq 0$, on the other hand, we have
\be
n\neq 0: \;\;\;\;\{\omega_J^{(-n)}\} = \{-\omega_J^{(n)}\}\;\;,
\ee
so that the $\pm n$-modes are paired. We will thus label the
frequencies $\omega_J^{(-n)}$ in such a way that
\be \la{kopi}
\omega_J^{(-n)} =
- \omega_J^{(n)} \ , \ \ \ \ \ \ \ \ J=1,...,2d \ .  \ee
Next,  we observe that the expression for the null eigenvectors we have found for
$d=2$ can be written as
\be
v_k = (-1)^k m_{1k}\;\;,
\ee
where $m_{ik}$ is the {\em minor} of $M_{ik}$, i.e.\  the determinant
of the matrix obtained from $M$ by removing the $i$'th row and $k$'th column. In
the $(2 \times 2)$ case, one  obviously has  $m_{11}=M_{22}$ and $m_{12}=M_{21}$.

As we show in Appendix A.1, this way of writing the null eigenvector generalises
to arbitrary dimensions. Thus the eigen-directions $a_{iJ}^{(n)}$ can for instance be
chosen to be
\be
a_{iJ}^{(n)} = (-1)^i m_{1i}(\omega_J^{(n)})\;\;,
\label{aiJ}
\ee
where  $m_{ij}(\omega^{(n)}_J)$ means the minor $m_{ij}$ of $M(\omega,n)$
evaluated for $\omega=\omega^{(n)}_J$. As in the case $d=2$ discussed
above, this is an acceptable choice of null eigenvector unless all
the $m_{1i}$ evaluated at $\omega_J^{(n)}$ are zero.  In that case,
instead of using the $m_{1i}$, one could of course also use the $m_{ji}$
for any other value of $j$. One can also use different $j$ for different
frequencies, and generically all these choices are equivalent.

Because the minor of a transpose is the transpose of the minor, for $n=0$
one can choose
\bea
a_{ij}^{+} &=& (-1)^i m_{1i}(\omega_j)\non
a_{ij}^{-} &=& (-1)^i m_{1i}(-\omega_j) = (-1)^i m_{i1}(\omega_j)\;\;.
\eea
Similarly, for $n\neq 0$, one can choose the eigenvectors for $-n$ to be given
by the transposes of the minors for $+n$.

Thus the solutions for the string modes are, explicitly,
\be
X^i_0 = (-1)^i\sum_{j=1}^{d}\
 [\xi^{+}_{j} m_{1i}(\omega_j)\ex{i\omega_j \tau}
 + \xi^{-}_{j} m_{i1}(\omega_j)\ex{-i\omega_j \tau}]\;\;,
\label{smz}
\ee
for $n=0$, and
\be
X^i_n = (-1)^i \sum_{J=1}^{2d} \xi^{(n)}_{J} m_{1i}(\omega_J^{(n)})
\ex{i\omega^{(n)}_J \tau}
\label{smn}
\ee
for $n \neq 0$.

%%%%%%%%%%%%%%%%%%%%%%%%%%%%%%%%%%%%%%%%%%%%%%%%
\subsection{The canonical commutation relations}

We now promote the $\xi$'s to operators and impose the canonical commutation
relations (CCRs)
\bea
&& [X^i(\sigma,\tau),X^k(\sigma',\tau)]=
[\Pi^i(\sigma,\tau),\Pi^k(\sigma',\tau)]=0\non
&& [X^i(\sigma,\tau),\Pi^k(\sigma',\tau)]=i\d^{ik}\delta
(\sigma-\sigma')\;\;,
\eea
where
\be
\Pi_i= \frac{1}{\pi}(\dot X_i  -  f_{ij} X^j)
\label{Pi}
\ee
are the canonical momenta  that follow  from
the string sigma model Lagrangian (\ref{Lag}) for the metric \rf{mou}.
Note that the the momenta $\Pi_i$ do not depend on
the null NS three-form field strength   \rf{dil}
in the \lc\ gauge.
It is clear that in order to get time-independent CCRs, the only non-zero
commutators can be between $\xi^+_j$ and $\xi^-_j$ (for $n=0$) and
$\xi^{(n)}_{J}$ and $\xi^{(-n)}_{J}$ for $n \neq 0$.
Let us call these
\bea
C_j &:=& [\xi^-_j,\xi^+_j] \non
C^{(n)}_J &:=& [\xi^{(-n)}_{J},\xi^{(n)}_{J}]\;\;.
\label{cjcj}
\eea
Note that obviously $C_J^{(-n)}=-C_J^{(n)}$.
We will show in Appendix B that imposing the canonical commutation
relations determines the $C_j$ and $C_{J}^{(n)}$ uniquely to be
\bea
C_{j} &=& \frac{1}{2 m_{11}(\omega_j)\
\omega_j\prod_{k\neq j}(\omega_j^2-\omega_k^2)}\non
C^{(n)}_{J} &=& \frac{1}{m_{11}(\omega^{(n)}_J)
\prod_{K\neq J}(\omega^{(n)}_J-\omega^{(n)}_K)}\;\;.
\label{cjn}
\eea
We now need to relate the oscillators $\xi_j^\pm$ with
\be
{}[\xi_j^-,\xi_k^+]=C_j\d_{jk}
\ee
to canonically normalised operators $\aa_j^\pm$ with
\be
{}[\aa_j^-,\aa_k^+]=\d_{jk}\ .
\label{apam}
\ee
Let us assume that the frequencies $\omega_j$ are {\it real} (and chosen to be positive).
Then the $C_j$ are also real. Reality of the string modes requires that
\be
(\xi_j^+)^{\dagger} = \xi_j^-\;\;.
\label{xirc}
\ee
If $C_j$ is positive, then one can define
\be
\aa_j^\pm = \xi_j^\pm/C_j^{1/2}\;\;\;\;\mbox{for}\;\; C_j > 0 \;\;,
\label{aj1}
\ee
with $(\aa_j^+)^{\dagger}=\aa_j^-$. If $C_j$ is negative, the situation is somewhat
different. One could try to define $\aa_j^\pm = \pm \xi_j^\pm/|C_j|^{1/2}$.
Then (\ref{apam}) is satisfied, but the $\aa_j^{\pm}$ are {\em not} the hermitian
conjugates of each other. Rather, one should
note that $C_j<0$ really means that
the creation operator $\aa_j^+$ is associated with the {\em negative} frequency
$(-\omega_j)$ and vice versa. In other words, one should define hermitian
conjugate operators by
\be
\aa_j^{\pm}= \xi_j^{\mp}/|C_j|^{1/2}\;\;\;\;\mbox{for}\;\; C_j < 0\;\;.
\label{aj2}
\ee

With the choice (\ref{aj1},\ref{aj2}), the bilinear $\xi_j^+\xi_j^- + \xi_j^-\xi_j^+$
is related to the number operator
\be
{\cal N}_j = \aa_j^+\aa_j^-
\ee
(with spectrum $n=0,1,2,\ldots$ on the Fock space defined by the vacuum annihilated
by $a_j^-$) by
\be
\trac{1}{2}(\xi_j^+\xi_j^- + \xi_j^-\xi_j^+) = |C_j|({\cal N}_j + \trac{1}{2})
\;\;.
\label{xixi1}
\ee
In the same way one can relate the $\xi_J^{(n)}$ to canonically normalised
oscillators $\aa_J^{(n)}$ with, say,
\be
{}[\aa_J^{(n)},\aa_K^{(m)}] = -\sign(n)\d_{n+m,0}\d_{JK}\;\;,
\ee
so that one has
\be
\trac{1}{2}(\xi_J^{(n)}\xi_J^{(-n)} +\xi_J^{(-n)}\xi_J^{(n)}) = |C_J^{(n)}|
({\cal N}_J^{(n)} + \trac{1}{2})
\label{xixi2}
\ee
with
\be
{\cal N}_J^{(n)} = \aa_J^{(n)}\aa_J^{(-n)}\;\;.
\ee
Hence, putting everything together one now has a completely explicit string
mode expansion. All the data entering the mode expansion (\ref{xfb}) in
this frequency basis for the solutions are determined by the matrix $M$ in
\rf{Mik}:
\begin{itemize}
\item
the frequencies are the roots of $M$
\item
the eigen-directions of the frequencies are
constructed from the minors of $M$
\item
the commutators of the $\xi$'s are determined
by the frequencies.
\end{itemize}

%%%%%%%%%%%%%%%%%%%%%%%%%%%%%%%%%%%%%%%%%%%%%%%%%%%%%%%%%%%%%%%%%
\subsection{The string Hamiltonian and its oscillator frequencies}

The light-cone Hamiltonian is\footnote{Note that written in terms of the
velocities $\dot{X}^k$ rather than the
momenta $\Pi^k$, the Hamiltonian does not
explicitly  depend  on $f_{ij}$. Recall also that we have set
$\a' = \ha, \ p_v=1$, and $H= - p_u$.
}
\be
H = \frac{1}{2\pi}\int_{0}^{\pi}d\sigma \ [\ \delta_{ij}(\dot{X}^{i}\dot{X}^{j}
+ X^{i\prime}X^{j\prime} - k_i X^{i} X^{j}) - 2 h_{ij} X^{i}X^{j\prime}\ ]
\ee
Here primes indicate $\sigma$-derivatives.
Inserting the mode expansion \rf{smz},\rf{smn}
for the $X^{i}$, the Hamiltonian
becomes a bilinear expression
in the oscillators $\xi_j^{\pm}$ and $\xi_{J}^{(\pm n)}$.
Since the light-cone Hamiltonian
for a smooth homogeneous plane wave in the stationary coordinates (\ref{mmm})
is
time-independent, it follows that the only non-zero terms are those proportional
to the diagonal combinations $\xi_j^{+}\xi_j^{-}$ or $\xi_{J}^{(n)}\xi_J^{(-n)}$.

The vanishing of the other contributions can be expressed as certain
identities about roots and minors. While these are guaranteed to hold,
by virtue of the time independence of the light-cone Hamiltonian, in low
dimensional cases one can also verify them explicitly.\footnote{It might
be desirable to have a general proof of these identities which does not
invoke reference to time-independence of the string Hamiltonian. A proof
of one of the required identities can be found in Appendix A.2.}

Because of the pairing of the $\pm n$ modes,  this Hamiltonian can be written as   a sum
\be\la{huu}
H = \sum_{n=0}^{\infty}H^{(n)} \;\;.
\ee
The zero-mode part or particle Hamiltonian is
\be
H^{(0)} = \frac{1}{2} \sum_{i=1}^d[(\dot{X}_0^i)^2 - \kk_i (X_0^i)^2]
\;\;.
\ee
Inserting the mode expansion (\ref{smz}) and
retaining only the terms proportional to $\xi_j^\pm\xi_j^\mp$, one finds
\be
H^{(0)} = \frac{1}{2}\sum_{j}
\sum_i (\omega_j^2 - \kk_i)m_{1i}(\omega_j) m_{i1}(\omega_j)
\ (\xi_j^+\xi_j^- + \xi_j^-\xi_j^+) \;\;.
\label{h0}
\ee
As the oscillators corresponding to different frequencies $\omega_j$ commute,
$H^{(0)}$ is a sum of  operators {\em without} cross interactions.
We see that in the frequency basis the Hamiltonian is automatically diagonal,
and there is no need to perform explicitly the unitary transformation to diagonal
form, as was done in section 2.6.

When the frequencies $\omega_j$ are real, the $\xi_j^\pm$ are related to
standard oscillators in the manner described in the previous section and
the above Hamiltonian is a sum of harmonic oscillators. On the other hand,
when the frequencies are, say, imaginary, then the reality conditions
require not (\ref{xirc}) but rather (assuming that the eigendirections
are real) $(\xi_j^\pm)^\dagger = \xi_j^\pm$. In this case the Hamiltonian
(\ref{h0}) also has the form of a harmonic oscillator one, but now with
imaginary frequencies, as can be seen by identifying the $\xi_j^\pm$ up to
constants with hermitian operators $p_j \pm q_j$. Similar considerations
apply in the case of complex frequencies.  From now on, unless stated
otherwise, we will assume that all frequencies are real.

Using (\ref{xixi1}) and the identity (\ref{mmmm}), one finds that
the Hamiltonian takes the form
\be\la{hze}
H^{(0)} = \sum^d_{j=1} \sign(C_j)\Omega_j ({\cal N}_j + \frac{1}{2})\;\;,
\ee
where
\be
\Omega_j = C_j m_{11}(\omega_j)\sum_i (\omega_j^2 - \kk_i)m_{ii}(\omega_j)\;\;.
\ee
Here we have split $|C_j| = \sign(C_j) C_j$, to keep track of the sign of the
spectrum -- as we had seen, a negative $C_j$ corresponds to an exchange of
positive and negative frequencies.

Using the expression for $C_j$ obtained in (\ref{cjsol}), one finds
\be
\Omega_j = \frac{\sum_i
(\omega_j^2 - \kk_i)m_{ii}(\omega_j)}{2\omega_j\prod_{k\neq j}(\omega_j^2-\omega_k^2)}
\;\;.
\ee
We now claim that, despite appearance,
the result for $\Omega_j $  is actually very simple, namely
that (up to $\sign(C_j)$) the oscillator frequencies of the  quantum string
Hamiltonian are equal to the frequencies of the classical string modes,
\be
\Omega_j = \omega_j\;\;,
\label{Oo}
\ee
This relation is easy to establish for four-dimensional plane waves,
i.e.\ $d=2$.  We thus have two frequencies $\omega_1$ and $\omega_2$,
and e.g.\ $\Omega_1$ is
\bea
\Omega_1&=&\frac{(\omega_1^2 -\kk_1)(\omega_1^2 + \kk_2)
+ (\omega_1^2 -\kk_2)(\omega_1^2 + \kk_1)}{2\omega_1(\omega_1^2-\omega_2^2)} \non
&=&
\frac{(\omega_1^4 - \kk_1 \kk_2)}{\omega_1(\omega_1^2-\omega_2^2)}
=
\frac{(\omega_1^4 - \omega_1^2\omega_2^2)}{\omega_1(\omega_1^2-\omega_2^2)}
= \omega_1\;\;,
\eea
where
\be
\kk_1 \kk_2 = \omega_1^2 \omega_2^2
\ee
follows from the fact that $\kk_1 \kk_2$ is the constant term in the
quadratic equation (\ref{ofour}) for $\omega^2$. Combined with the
explicit expression (\ref{fbf}) for the classical frequencies, this
reproduces the result of section 2.6 based on explicit diagonalisation of
the Hamiltonian by a unitary transformation.

General validity of (\ref{Oo}) is equivalent to the identity
\be
\sum_i (\omega_j^2 - \kk_i)m_{ii}(\omega_j) = 2 \omega_j^2 \prod_{k\neq j}(\omega_j^2-
\omega_k^2)\;\;.
\ee
Using the same kind of manipulations as in the case $d=2$, it is straightforward
to verify this identity directly for $d=3$, and we certainly expect it to be true in
general. If this is the case, then this ought to be obvious on a priori
grounds.\footnote{This may be more apparent after doing this calculation in  phase space
variables.} Note, however, that neither from the present point of view nor from
the analysis of section 2.6 this is completely manifest.  This is a
reflection of the non-diagonal nature of the original problem.

Likewise, for the string ($n\not=0$) modes, the
Hamiltonian takes the form
\be\la{han}
H^{(n)}=\sum^{2d}_{J=1}\sign(C_J^{(n)})
\Omega_J^{(n)}({\cal N}_J^{(n)}+\frac{1}{2})\;\;,
\ee
where $\Omega_J^{(n)}$ is the sum of two terms -- one coming from the metric,
the other from the $B$-field,
\be
 \Omega_J^{(n)} = C_J^{(n)}m_{11}(\omega_J^{(n)})
   \sum_{i,j} [((\omega_J^{(n)})^2 - \kk_i + 4n^2)\d_{ij}
   - 4 i n (-1)^{i+j} h_{ij}]m_{ij}(\omega_J^{(n)}) \ .
   \label{hfre}
\ee
The above expression for the frequencies can be simplified somewhat. Using
\rf{idd1},  \rf{hfre} can be rewritten as
\be
\Omega_J^{(n)}=2 \omega_J^{(n)} C_J^{(n)}m_{11}(\omega_J^{(n)})
   \sum_{i,j} [\omega_J^{(n)} \d_{ij}
   + i  (-1)^{i+j} f_{ij}]m_{ij}(\omega_J^{(n)}) \ .
   \ee
As in the particle case (\ref{Oo}), we expect
the frequencies $\Omega_J^{(n)}$ to be exactly equal to the string mode frequencies,
\be\la{bef}
\Omega_J^{(n)}= \omega_J^{(n)}\;\;.
\label{freeq}
\ee
General validity of (\ref{freeq}) is equivalent to the identity
\be
2 \sum_{i,j} [\omega_J^{(n)} \d_{ij}
   + i  (-1)^{i+j} f_{ij}]m_{ij}(\omega_J^{(n)})=\prod_{K\not=J}
(\omega_J^{(n)}-\omega_K^{(n)})~
   \label{conid}
\ee
which can be verified for $d=2$ after some computations which are explained
in section 4.4.

%%%%%%%%%%%%%%%%%%%%%%%%%%%%%%%%%%%%%%%%%%%%%%%%%%%%%%%%%%%%%%%%%%%%

\subsection{The level matching condition}
%%%%%%%%%%%%%%%%%%%%%%%%%%%%%%%%%%%%%%%%%%%%%%%%%%%%%%%%%%%

In general, the \lc\ Hamiltonian  should be supplemented by the
condition of translational invariance along the closed string,
i.e.
\be
\int^\pi_0  d \s \ \Pi_i X'^i =0 \ ,
\ee
which should be imposed on the string spectrum.
Using the
definition (\ref{Pi}) of the momenta $\Pi_i$ and the mode expansion
(\ref{smn}), one finds that more explicitly this constraint takes the form
\be
-2i\pi
\sum_{n>0}
n(:\Pi^i_{n}X^{i}_{-n} - \Pi^{i}_{-n}X^{i}_{n}:) = 0\;\;,
\ee
where $:\ :$ refers to the normal ordering prescription of the previous
section and
\be
\Pi^i_n= \frac{(-1)^i}{\pi}\sum_{J=1}^{2d}
\xi_J^{(n)}[i\omega_J^{(n)}\d_{ik} - (-1)^{i+k}f_{ik}]
m_{1k}(\omega_J^{(n)})\ex{i\omega_J^{(n)}\tau}\;\;.
\ee
The
constraint is time-independent, as can be checked by using the
equations of motion (\ref{ddx}). Hence we can drop all the
non-diagonal terms $\xi_J^{(n)} \xi_{K}^{(-n)}$, $J \neq K$, in
the expansion of the constraint as these would enter with
time-dependent phases.\footnote{Once again, vanishing of these contributions
can be rephrased as certain identities about roots and minors.} 
After some rearrangement, using the
identity (\ref{mmmm}), we can write the level-matching constraint
in terms of the number operators ${\cal N}_J^{(n)}$ as
\be
\sum_{n>0}2n \sum_{J=1}^{2d}m_{11}(\omega_J^{(n)}) |C_{J}^{(n)}|
\sum_{j,k=1}^d[\omega_J^{(n)}\d_{jk}+i(-1)^{j+k}f_{jk}]
m_{jk}(\omega_J^{(n)}) {\cal N}_J^{(n)}=0\;\;.
\ee
Using the
explicit expression (\ref{CJs}) for the $C_J^{(n)}$, we can
rewrite this condition as
\be \sum_{n>0} n\sum_{J=1}^{2d} S_J^{(n)}
{\cal N}_J^{(n)} = 0 \label{lmc}
\ee
where
\be
S_J^{(n)} =
2\sign(m_{11}(\omega_J^{(n)}))
\frac{\sum_{j,k=1}^d[\omega_J^{(n)}\d_{jk}+i(-1)^{j+k}f_{jk}]
m_{jk}(\omega_J^{(n)})}{|\prod_{K\neq
J}(\omega_J^{(n)}-\omega_K^{(n)})|}\;\;. \label{sjn}
\ee
Assuming the general validity of \rf{conid}, we find that
\be
S_J^{(n)} = \sign(m_{11}(\omega_J^{(n)}))  \sign (\prod_{K\neq
J}(\omega_J^{(n)}-\omega_K^{(n)}))
\label{sjn2}
\ee
and so
\be
S_J^{(n)}=\pm 1\;\;.
\ee
Comparison of (\ref{sjn2}) and the expression for $C_J^{(n)}$ (\ref{cjn})
shows that
\be
S_J^{(n)} = \sign(C_J^{(n)})\;\;.
\ee
This means that the signs appearing in the string oscillator spectrum agree with
the signs in the level-matching condition.

In the four-dimensional case we will show in section 4.4
that typically (but not necessarily), and in particular for large $n$,
we have
\be
S_J^{(n)}=\sign(\omega_J^{(n)})\;\;,
\ee
as might have naively been expected for the level-matching condition.
In particular, therefore, for large $n$ the string Hamiltonian is a sum
of harmonic oscillators with positive frequencies
\be
\sign(C_J^{(n)})\omega_J^{(n)}=|\omega_J^{(n)}|\;\;.
\ee
However for small $n$ (small compared to  $f_{ij}$, measured in units of $p_v$),
there may be deviations from this and both the Hamiltonian and the level-matching
conditions may involve some unusual signs.

%%%%%%%%%%%%%%%%%%%%%%%%%%%%%%%%%%%%%%%%%%%%%%%%%%%%%%%%%%
\section{Aspects of the string spectrum for four-dimensional plane waves}

In the previous sections we have shown that, generically, the particle
Hamiltonian in a homogeneous plane wave background takes the form of
a set of decoupled harmonic oscillators, with real or possibly
complex frequencies.  In the following, we will analyse
various aspects of the resulting particle and string spectra in the case
of four-dimensional plane waves.

%%%%%%%%%%%%%%%%%%%%%%%%%%%%%%%%%%%%
\subsection{Analysis of the particle spectrum}

In the particle case we found that  the frequencies, masses and oscillator
commutators \rf{cjcj} are (see \rf{freq},\rf{fbf} for the frequencies)
\be\la{daan}
\omega_{1,2}^2 = \frac{1}{2}(4\ff^2-\kk_1-\kk_2 \mp \D)\ ,
\ \ \ \ \ \   \D^2=(4\ff^2-\kk_1-\kk_2)^2 - 4 \kk_1\kk_2  \ , \ee
\be
m_{1,2} =  \frac{2\D}{\D+\kk_1-\kk_2\mp 4\ff^2}\  , \ \ \ \ \ \ \
C_{1,2} = \mp\frac{1}{2\omega_{1,2}(\omega_{1,2}^2+\kk_2)\D}\;\;.
\ee
We see that when $\D = 0$, so that $\omega_1^2 = \omega_2^2$,
the masses $m_i$ vanish and correspondingly the $C_i$ are infinite.
This can happen only if $\kk_1$ and $\kk_2$ are both positive and $
\sqrt {\kk_1} \pm  \sqrt{\kk_2} = 2 \ff$.  In that  case  the unitary
transformation  \rf{uni} which related the original system to the system
of two decoupled  harmonic oscillators breaks down because the phase
$\theta_2$ becomes infinite.
One of the $C_i$ will also be infinite when one of the frequencies is
zero. This implies that one of the $\kk_i$ is zero and that one of the
masses $m_i$ is infinite.

When there is no rotation, $\ff = 0$, the frequencies are $\omega_i^2 =
-\kk_i$, with masses $m_i=1$. We see that in this case one of the $C_i$
would blow up, but this is due to the fact that
the minors $m_{11}=m_{12}=0$ so that
this is simply a reflection of a bad choice of a  null eigen-direction, which
can be fixed by choosing different null vectors
for different frequencies (see the discussion following (\ref{4dai})).

Assuming that both frequencies $\om_i$ are non-zero and  real (we
shall always assume that if the $\om_i$ are real they are chosen to
be  positive), the Hilbert space of the theory is that of two harmonic
oscillators with generically different frequencies.  The spectrum of
the Hamiltonian $H^{(0)}$ is given by the (almost) standard oscillator
expression
\be
\la{spee}
\E= \pm\omega_1 (n_1+\ha)\pm\omega_2 (n_2+\ha)  \ ,
\ee
where $n_1,n_2=0,1,2,...$. As explained in section 2.6 and section 3.6,
the sign of the spectrum depends on the sign of the $m_i$ or $C_i$. We
will come back to this issue below.

One finds that   $\om^2_i$  are positive if, e.g.,  $\kk_i < 0$
(then $0 < \om^2_1 < \om^2_2$).  If $\kk_i >0$, then the frequencies
$\om^2_i$ are  positive  provided $ 4\ff^2 > \kk_1 + \kk_2$.
If the signs of $\kk_1$ and $\kk_2$ are different, then
$\om^2_1 < 0,\  \om^2_2 > 0$  and
thus the  oscillator system is unstable in
 one direction.

Let us now take into account the conditions (see section 2.3)
on the parameters
$\kk_i$ and $\ff$
imposed by the conformal invariance conditions (i.e. by the Einstein equations).
First, in  the pure-metric case
(constant dilaton and zero 3-form strength)
with vanishing rotation parameter $\ff=0$,
the Ricci-flatness implies that
\be
\la{rel}
 \kk_1+ \kk_2 =0 \ , \ \ \ \ {\rm i.e.} \ \ \ \ \  \D= \kk_1 - \kk_2  \ , \ \ \
  \om^2_1= - \kk_1   \ ,  \   \ \   \om^2_2 = - \kk_2  \ , \ \ \ \   m_{1,2}=1 \ .  \ee
Thus  one of the  frequencies  is
necessarily imaginary, and we  get  an inverted
oscillator problem in one of the two directions: the
absence of a  stable ground state  just means that
the particle  ``escapes'' to infinity (the classical
geodesics are  also  ``pushed'' to infinity).\foot{
 As in the standard plane wave cases
 discussed in \ci{marol},  the negative values of the \lc\ gauge
 energies  do not of course imply any real instability
 of the plane wave background  but are rather related to
  the fact that
 some geodesics may escape to infinity.}

For metrics with
$\ff\not=0$ the Ricci-flatness condition  implies that (cf. \rf{coon})
\be \la{pul}
 \kk_1+ \kk_2 = 2 \ff^2 \ , \ \ \
{\rm i.e.} \ \ \ \ \ \  \D= \kk_1 - \kk_2
 \ ,\ \ \  \ \ \ \
\om^2_{1}   = \kk_2   \ ,  \ \ \
\om^2_{2}   = \kk_1  \ , \ee
\be
   m_{1} = -{ \kk_1 - \kk_2  \ov 2 \kk_2 }
\ ,\ \ \ \ \ \ \ \   m_2 =
 { \kk_1 - \kk_2  \ov 2\kk_1 }
  \ .
\ee
Compared to \rf{rel}  here    $\om_1^2$ and $\om^2_2$
are multiplied by -1  and their absolute values are interchanged.
It appears as if the  role of the rotation parameter $\ff$ is thus to
{\it reverse
the signs} of the two frequencies compared to the $\ff=0$ case, i.e.
one is tempted to
conclude that (like in the case of
non-Ricci-flat  plane waves with extra fluxes)
the  Ricci-flat plane waves with {\it non-zero}
rotation parameter
$\ff$  may have  {\it  localised}  point-like particle dynamics
in transverse directions
(both at the classical and quantum level):
if  $\kk_1$ and $\kk_2$ are  positive,
the particle is described by an effective Hamiltonian  of
two decoupled oscillators with {\it real} frequencies.

For generic non-vanishing $\kk_i > 0$
one of the two masses in \rf{hmm} is negative.
That means that one of the effective oscillator
Hamiltonians $H_i$ enters their sum  with the negative sign,
and thus one of the two oscillator  energy spectrum terms in
\rf{spee}  should be taken with  the negative sign.
This is not surprising given that for positive $\kk_1$ and $\kk_2$
the original Hamiltonian \rf{hah} is not positive definite, and
can also be understood from the fact that $m_i <0$
is possible iff $C_i<0$,
so that the role of positive and negative frequencies is interchanged (see section
3.5). Thus we  are in a novel situation  where the
\lc\  energy spectrum is discrete
but not positive,
\be\la{spen}
\E= - \omega_1 (n_1+\ha) +\omega_2 (n_2+\ha) \ .
\ee

%%%%%%%%%%%%%%%%%%%%%%%%%%%%%%%%%%%%%%%%%%%
As it is clear from  \rf{sep},\rf{coon}, the role of the  3-form or
the  $\hh$-parameter is to effectively reduce the value
 of $\ff$: for non-zero $\hh$ we have
  $\kk_1 + \kk_2 = 2( \ff^2 - \hh^2)$ and thus
\be
\om^2_{1,2} = \ff^2 + \hh^2  \mp \ha \D  \ ,
\ \ \ \ \ \ \ \ \ \   \D^2 = (\kk_1 - \kk_2)^2  + 16 \ff^2 \hh^2 \ .
\ee
Depending on the value of
 $\kk_1 - \kk_2$  both  frequencies may be  real or one of them
may be imaginary. In particular, in
the chiral null model case
with an additional $SO(2)$ symmetry in $x_i$-directions,
i.e. for    $\ff^2 = \hh^2$ and  $\kk_1= \kk_2$,
we get  $\om^2_{1} = 0, \ \om_2^2 = 4 \ff^2, \  m_1=\infty,\  m_2 =1 $,  i.e.
we get again the Landau-type  spectrum, in agreement with
\ci{ruts}.

In the case of a non-trivial  ($u$-dependent) dilatonic background,
the dilaton can be eliminated from the Klein-Gordon
 equation
by a redefinition $\Psi \to e^{\p(u)} \Psi$
of the field  in \rf{kg} (see \ci{prt} for details). For a non-constant  dilaton
given by the  case I of \rf{dio} the positivity \rf{ooi}
of the parameter $\m$ in \rf{dim} or \rf{kop}
implies   the restriction $ \kk_1 + \kk_2 < 2 (\ff^2-\hh^2)$.
Interestingly, the  latter
condition ensures also the reality of $\D$ in \rf{daan}.

%%%%%%%%%%%%%%%%%%%%%%%%%%%%%%%%%%%%%%%%%%%%%%%%
\subsection{The anti-Mach metric}

A border-line case is when one of the  parameters $\kk_1$ or $\kk_2$
vanishes. This corresponds to the anti-Mach metric  discussed in \ci{mm}.
If we choose,   as  in \rf{mah},
$\kk_1= 2 \ff^2, \  \kk_2 = 0$,
we get
\be
\la{rees}
 \omega^2_{1} = 0\ , \ \ \ \ \ \ \ \ \omega^2_2= 2 \ff^2 \  ,
\ \ \ \ \ \   m_1 = -\infty, \ m_2 = \ha  \ . \ee
In this case  the phases  \rf{tat} are
$\theta_1=0, \ \theta_2= -\ff^{-1}$.

The vanishing of one of the frequencies suggests that the spectrum
 should contain a discrete oscillator contribution  in one
direction and a continuous  free-particle contribution in another,
in agreement  with the previous analysis of \ci{mm}.
However, since here  one of the effective masses is infinite,  this
case needs special consideration.

In \cite{mm} this conclusion was reached by noting that the
coordinate transformation
\be
v \ra v + 2\ff xy
\ee
puts the anti-Mach metric (which corresponds to the choice of the parameters
(\ref{mah}))
\be
ds^2 = 2 dudv + 2 \ff^2 x^2 du^2 + 2 \ff (x dy - ydx) du + dx^2 + dy^2
\ee
into the form
\be
ds^2 = 2 du dv + 2 \ff^2 x^2 du^2  + 4 \ff x dy du + dx^2 + dy^2\;\;.
\label{aMm}
\ee
In this coordinate system, there is clearly a translation invariance in the
$y$-direction which gives rise to a continuous contribution to the spectrum.

This coordinate transformation can be equivalently regarded
as a gauge transformation
of the gauge field $A_i$ in \rf{met}. Thus, to make the discussion of the $\kk_2=0$ case
more transparent,  it is useful to follow
the analysis  of the Landau problem in
a different gauge for $A_i$ than used in  \rf{jj},\rf{one},\rf{od}, namely,
\be A_x = 0\  ,\ \ \ \ \ \ \  \ A_y = 2 \ff x \ .  \ee
Then   the Hamiltonian \rf{hmmm} takes the following form
(cf. \rf{hah})
\be \la{hih}
H=-\ha [\partial_x^2+(\partial_y-2i \ff  x)^2 +\kk_1 x^2 ] \ .
\ee
The case $\kk_1=0$ corresponds to the standard Landau problem,
and the case of $\kk_1= 2 \ff^2$ to the anti-Mach metric  case.
Since $y\to y + c$ is a symmetry,
we may  use the Fourier transformation in $y$,
i.e. $ \partial_y \to i p_y$.  Then
\be \la{hoh}
H=-\ha \partial_x^2  + V  \ , \ \ \ \ \ \
V = \ha [ (2 \ff  x - p_y)^2  -  \kk_1 x^2 ] \ .
\ee
The potential here is positive at large $x$ both in the Landau model case $\kk_1=0$
and in the anti-Mach case $\kk_1 = 2 \ff^2$.
In the Landau model case  we can  ``absorb'' $p_y$ into $x$  by a constant shift
and thus the resulting spectrum is the discrete spectrum of a one-dimensional
harmonic oscillator with  mass 1 and frequency $\om_1= 2  \ff$.
In the anti-Mach case we get, diagonalising the quadratic form
in the potential,
\be \la{popi}
V = \ha [ (2 \ff  x - p_y)^2  -  \kk_1 x^2 ]
=  (\ff  x - p_y)^2  -  \ha p_y^2  \  .
\ee
The resulting analog of the spectrum  \rf{spee}
is indeed continuous but is {\it not}  positive definite (cf.\ \rf{spee}):
\be\la{sos}
 \E= \omega_2 (n_2+\ha) - \ha p^2_y  \ ,  \ \ \ \ \ \ \ \ \  \
\om_2=  \sqrt{2} \ \ff \ . \   \ee
This is, indeed, in agreement  with the spectrum found in \ci{mm}.
As expected on the basis of  one of the two  effective masses going
to minus infinity as the frequency goes to zero in \rf{rees}, here the
free-motion contribution enters with negative sign relative
to the mass term.
This is the opposite to what happens in the flat space limit where
(for
$p_v=1$) one has
$ H=-p_u= \ha p^2_x + \ha p^2_y $. This result
 is  obviously a reflection of an
inherent ``unboundedness''  of trajectories in this
 case -- the \lc\ energy
can take any possible negative values.

It is clear now that similar features are shared by the general case
of non-zero $\kk_2$ where  again one of the masses is negative: there,
as explained above (see \rf{spen}),  the negative contribution to
 the \lc\
spectrum is not continuous as in the anti-Mach case but discrete.

%%%%%%%%%%%%%%%%%%%%%%%%%%%%%%%%%%%%%%%%%%%%
\subsection{Analysis of the string spectrum}

The string mode frequencies are the roots of the equation (\ref{4droots}),
\be
\omega^4 + (\kk_1 + \kk_2 - 4\ff^2 - 8 n^2) \omega^2  - 16n \ff\hh \omega
+ (\kk_1-4n^2)(\kk_2-4n^2)
- 16 n^2 \hh^2 = 0\;\;.
\label{4droots2}
\ee
In general the roots $\omega_{J}^{(n)}$, $J=1,2,3,4$, of this equation satisfy
the identities
\bea
\sum_{J=1}^4 \omega_J^{(n)}&=&0\;\;, \label{lomid}\\
\sum_{J<K} \omega^{(n)}_J\omega^{(n)}_K &=& \kk_1 + \kk_2 - 4\ff^2 - 8 n^2\ , 
\label{qomid}\\
\sum_{J<K<L}\omega^{(n)}_J\omega^{(n)}_K\omega^{(n)}_L &=& 16 n \ff\hh\ , 
\label{comid}\\
\omega^{(n)}_1 \omega^{(n)}_2 \omega^{(n)}_3 \omega^{(n)}_4&=&(\kk_1-4n^2)(\kk_2-4n^2)
- 16 n^2 \hh^2
\;\;.
\label{omid}
\eea
Note that for large $n$, the product of the roots is positive. Since the sum of the
roots is zero, this means that for sufficiently large $n$ there will be two
positive and two negative frequencies.

When $\hh =0$, eq. \rf{4droots2} is a quadratic equation for $\omega^2$ and the
frequencies \rf{fern} are
\be
\omega^{(n)2}_{{1,2}} = \frac{1}{2}(4\ff^2
+ 8 n^2 -\kk_1-\kk_2 \pm \D_n)\ ,
\ee
where
\bea
\D_n^2&=& (4\ff^2 + 8 n^2 -\kk_1-\kk_2)^2 - 4 (\kk_1-4n^2)(\kk_2-4n^2)
\non
&=&
(\kk_1-\kk_2)^2-8(\kk_1 + \kk_2)\ff^2 + 16 \ff^4 + 64 \ff^2 n^2\;\;.
\eea
We see that, regardless of the values of $\kk_i$ and $\ff$, the
frequencies are real for large enough values of $n$,
i.e. both the ``rotation'' or ``magnetic'' parameter $\ff$
{\it  and} the string  excitation level $n$  work towards
stabilising the string motion.

A simple example is $\kk_1=\kk_2=\ff^2$, which satisfies the Ricci flatness
condition (\ref{coon}). In this case, $\D_n^2 = 64 \ff^2 n^2$, and the four
distinct real frequencies are
\be
\omega_J^{(n)}=\pm\ff \pm 2n\;\;.
\label{exfkk}
\ee

An interesting possibility in the case when the particle motion is unstable
is to get new string states that have (nearly) zero frequencies,
i.e.\ new light states.
For example, in the case of the anti-Mach metric (\ref{aMm}) with
parameters (\ref{mah}), $\kk_1= 2 \ff^2, \kk_2 = 0$, we find
\be
\omega^{(n)2}_{{1,2}} = \ff^2 + 4n^2 \pm
\sqrt{\ff^4 + 16 \ff^2 n^2}\ .
\label{anti}
\ee
For the zero mode ($n=0$)  we reproduce the result \rf{rees} \ci{mm}
of the point-particle analysis that one of the two frequencies vanishes.
Clearly,  $\omega_1^{(n)}$ is always real.
$\omega_2^{(n)}$, on the other hand, can be real, zero, or imaginary.
In fact, for the special values $\ff = \sqrt 2 n$, the frequency vanishes,
and one has a new massless state. For $\ff > \sqrt 2 n$ the frequency is
imaginary, i.e.\ for fixed $\ff$ for sufficiently large $n$ all
frequencies are real and the motion is oscillatory.

Note that  compared to the  homogeneous  R-R plane wave  model  in
\ci{mt} where the spectrum had the form  $\omega^{(n)}= \sqrt{|\kk|  +   4
n^2}$  (with $\kk$ determined by the R-R flux)
here we have  more intricate (``double square root'')
dependence on the oscillator level $n$.  Similar string oscillator
spectra were found previously in the case of ``standard'' (no rotation)
plane waves supported by a null  (and non-constant,
$\sim e^{\mu u}$)\  $B_{MN}$  background in \ci{war} (see also 
\cite{casero}). One may thus contemplate possible
connection to deformations of conformal gauge theories via Penrose limits
of the corresponding geometries (cf.\ \ci{prez}).

When $\ff = 0$, the parameter  $\hh$ appears only in the constant term and one has
\be
\omega^{(n)2}_{J} = \frac{1}{2}(8 n^2 -\kk_1-\kk_2) \pm
\frac{1}{2}\sqrt{(\kk_1-\kk_2)^2 + 64 n^2 \hh^2}\;\;.
\ee
In particular, when $\kk_1 = \kk_2 \equiv \kk$ and $\kk = - \hh^2$
(so that \rf{coon} is satisfied for
 constant dilaton), one can take  the square root
 and the frequencies are (cf.\ (\ref{exfkk}))
\be\la{hop}
\omega^{(n)}_{J} =  \pm (2n \pm \hh)\;\;.
\ee
Here for special values of $\hh$ one finds additional
states with zero \lc\ energy.
This case (related by a coordinate transformation
to a WZW \ci{NW} or chiral null model case)
was considered previously in \ci{ruts,bmn,rt}.

In general, when $\ff\neq 0$ and $\hh \neq 0$, the frequencies are complicated
functions of the parameters. We can write the defining equation (\ref{4droots2})
as
\be\la{jok}
(\omega^2 - 4 n^2 + \kk_1)(\omega^2 - 4 n^2 + \kk_2)=
4(\ff \omega + 2n \hh)^2
\;\;.
\ee
Eq. \rf{jok} simplifies in the case $\kk_1=\kk_2=\kk$,
when the four roots are the solutions of the
two quadratic equations
\be
\omega^2 - 4 n^2 + \kk = \pm (2\ff\omega + 4 n\hh)\;\;,
\ee
namely
\bea
\omega_{1,2} &=& \ff \pm \sqrt{\ff^2 + 4 n^2 - \kk + 4 \hh n}\non
\omega_{3,4} &=& -\ff \pm \sqrt{\ff^2 + 4 n^2 - \kk - 4 \hh n}
\eea
Let us now consider the chiral null
model case $\ff = \hh$ (cf.\ \ci{ruts,rt}).
With $\ff = \hh$ and $\kk_1 = \kk_2$, the
conformal invariance condition \rf{coon}  $\kk_1 + \kk_2
= 2\ff^2 - 2\hh^2$ implies that
 $\kk_1=\kk_2 =0$, so that the four frequencies are thus
simply (cf. \rf{hop})
\be
\omega_{1,4}^{(n)}=2(n\pm\ff)\;\;,\;\;\;\;\;\;\omega_2^{(n)}=\omega_3^{(n)}=-2n \;\;.
\ee
 Note  that the matrix $M(\omega,n)$
in \rf{mam}
evaluated at
$\omega_2^{(n)}=\omega_3^{(n)}$ is identically zero.
Therefore,  it has
two linearly independent null eigen-directions, and so
the frequency base ansatz
(\ref{xfb}) provides us with  the full set of four linearly independent solutions,
even though two of the frequencies are equal (and the construction of the null
eigendirections in terms of the minors of $M$ is not applicable).

%%%%%%%%%%%%%%%%%%%%%%%%%%%%%%%%%%%%%%%%%%%%%%%%%%%%%%%%%%%%%%%
\subsection{The level matching condition}
%%%%%%%%%%%%%%%%%%%%%%%%%%%%%%%%%%%%%%%%%%%%%%%%%%%%

In section 3.7 we had derived the general level matching condition (\ref{lmc}),
\be\la{kopp}
\sum_{n>0} n\sum_{J=1}^{2d} S_J^{(n)} {\cal N}_J^{(n)} = 0
\ee
where
\be
S_J^{(n)} =
2\sign(m_{11}(\omega_J^{(n)}))
\frac{\sum_{j,k=1}^d[\omega_J^{(n)}\d_{jk}+i(-1)^{j+k}f_{jk}]
m_{jk}(\omega_J^{(n)})}{|\prod_{K\neq J}(\omega_J^{(n)}-\omega_K^{(n)})|}\;\;.
\ee
To evaluate this in the present case of four-dimensional plane waves,
it is cumbersome (and unnecessary) to use the explicit expressions for the
roots. Rather, we will use the general identities (\ref{lomid}-\ref{omid}) 
satisfied by the
roots of any quartic equation (with zero cubic term).

It follows from the form of the matrix $M$ (\ref{mam}) that
\be
\sum_{j,k=1}^d[\omega_J^{(n)}\d_{jk}+i(-1)^{j+k}f_{jk}] m_{jk}(\omega_J^{(n)})
= \omega_{J}^{(n)}(2\omega_{J}^{(n)2} - 8 n^2 + \kk_1 + \kk_2 - 4 \ff^2) - 8n\hh\ff
\;\;.
\label{sjn1}
\ee
Using the quadratic and cubic identities (\ref{qomid},\ref{comid}), 
we can write this as
\be
(\ref{sjn1})=
\omega_{J}^{(n)}(2\omega_{J}^{(n)2}+\sum_{K<L}\omega^{(n)}_K\omega^{(n)}_L)
-\frac{1}{2}\sum_{K<L<M}\omega_{K}^{(n)}\omega_{L}^{(n)}\omega_{M}^{(n)}\;\;.
\ee
Using the linear identity (\ref{lomid}), we can write
\be
2\omega_{J}^{(n)2}=\frac{1}{2} \omega_{J}^{(n)2}-\frac{3}{2}\omega_{J}^{(n)}
\sum_{K\neq J} \omega_{K}^{(n)}\;\;.
\ee
Then one finds that the r.h.s. of (\ref{sjn1}) becomes
\be
\frac{1}{2}\prod_{K\neq J}(\omega_{J}^{(n)}-\omega_{K}^{(n)})\;\;.
\ee
This has two immediate consequences. First of all, it establishes the 
identity (\ref{conid}) ensuring that, in the four-dimensional case, 
we indeed have equality of the string mode and oscillator frequencies,
$\omega_J^{(n)}=\Omega_J^{(n)}$. Furthermore, the same identity now 
implies (as anticipated in section 3.7) that
\be
S_J^{(n)}=\sign(m_{11}(\omega_{J}^{(n)}))\;\sign(\prod_{K\neq
J}(\omega_{J}^{(n)}-\omega_{K}^{(n)})) = \pm 1\;\;.
\ee
For example, in the case $\ff=0$, it is now easy to see that
\be
S_J^{(n)}=\sign(\omega_{J}^{(n)})\;\;,
\ee
so that \rf{kopp} becomes the same as
the standard flat space level-matching condition.

To show that this is not necessarily the case when $\ff\neq 0$, it suffices
to consider the example $\kk_1=\kk_2=\ff^2$, $\hh=0$, with spectrum
$\omega_J^{(n)}=\pm \ff\pm 2n$ (\ref{exfkk}). In this case one finds
\be
S_J^{(n)}=\sign(\pm 2n)\;\;.
\ee
Evidently this agrees with $\sign(\omega_J^{(n)})$ for sufficiently large values
of $n$, $2n > \ff > 0$, but deviates from this for $2n < \ff$ (recall that we
have set $p_v=1$ -- in general the condition is $2n > p_v\ff$, which for any
$n$ can be violated for sufficiently large $p_v$).

To show this more generally  consider the case  where two of the
frequencies are positive $\omega^{(n)}_{1,2}$ and the other two $\omega^{(n)}_{3,4}$
are negative. We also assume that $-k_2+4n^2>0$ (this reduces to the inequality
$2n > \ff$ in the above example).  We order  the frequencies as
$\omega^{(n)}_1>\omega^{(n)}_2>\omega^{(n)}_3>\omega^{(n)}_4$. The polynomial $P(\omega)$
which has the frequencies as roots \rf{4droots2} is positive for $|\omega|\rightarrow
\infty$. In addition as it can be seen from \rf{jok}, it takes
negative values at $\omega^2=-\kk_2+4n^2$ and $\omega^2=-\kk_1+4n^2$.
 Therefore there is at least one root between $+\infty$ and $\omega=\sqrt{-\kk_2+4n^2}$.
 Since $\omega^{(n)}_1$ is the largest root,
 \be
 \prod_{J\not= 1} (\omega_1^{(n)}-\omega_J^{(n)})>0
 \ee
 and
 \be
\omega_1^{(n)}>\sqrt{-\kk_2+4n^2}~,
\ee
which imply that $S_1^{(n)}=\sign (\omega_1^{(n)})=1$.
Similarly, it can be shown that $S_4^{(n)}=\sign (\omega_4^{(n)})=-1$
by observing that $\omega_4^{(n)}<-\sqrt{-\kk_2+4n^2}$ and so
$(\omega_4^{(n)})^2>-\kk_2+4n^2$.
It remains to show the statement for the other two frequencies
$\omega^{(n)}_2$ and $\omega^{(n)}_3$. There are two possibilities,
either $\omega^{(n)}_2>\sqrt{-\kk_2+4n^2}$ or $\omega^{(n)}_2<\sqrt{-\kk_2+4n^2}$.
We exclude the former case. If $\omega^{(n)}_2>\sqrt{-\kk_2+4n^2}$, then
$P(\omega)$ has to have another positive root which is larger than $\sqrt{-\kk_2+4n^2}$.
If it did not, the value of $P(\omega)$ at $\sqrt{-\kk_2+4n^2}$ would have
been positive.
However this is not the case as we have shown. But we have two positive and two
negative frequencies, so we have to take $\omega^{(n)}_2<\sqrt{-\kk_2+4n^2}$.
In this case, it is easy to see that
$S_2^{(n)}=\sign (\omega_2^{(n)})=1$. Similarly, we can show that
$S_3^{(n)}=\sign (\omega_3^{(n)})=-1$.

When there are three positive frequencies and one negative, the formula
$S_J^{(n)}=\sign(\omega_J^{(n)})$ is not valid as can be seen by an
argument similar to the above. However, since we have seen that for
sufficiently large values of $n$ there are precisely two positive and
two negative frequencies, this establishes that typically, and for large $n$,
we indeed have $S_J^{(n)}=\sign(\omega_J^{(n)})$.

%%%%%%%%%%%%%%%%%%%%%%%%%%%%%%%%%%%%%%%%%%%%%%%%%%
\subsection{The zero-point energy}

It follows from (\ref{hze},\ref{han}) that the zero-point energy or
normal ordering constant $\ep_0$ has the form
\be
\la{noo}
\ep_0 = \frac{1}{2}\sum_{j=1}^d \sign(C_j)\omega_j
+ \sum_{n=1}^{\infty}s_n\;\;,
\ee
where
\be
s_n= \frac{1}{2}\sum_{J=1}^{2d}\sign(C^{(n)}_J)\omega_J^{(n)} - 2dn\;\;.
\ee
In  $s_n$  we have included the contribution  $(-2dn)$
of  the superstring fermions  which are decoupled from
the background in the \lc\ gauge (see section 2.5).

We had already seen that, for large $n$, $\sign(C^{(n)}_J)\omega_J^{(n)}=
|\omega_J^{(n)}|$, so that for sufficiently large oscillator number the
contribution to the zero-point energy is given by the sum of the absolute
values of the frequencies,
\be
s_n = \frac{1}{2}\sum_{J=1}^{2d}|\omega_J^{(n)}| - 2dn\;\;.
\ee
In particular, for $d=2$ and $\ff=0$ or $\hh =0$ the four frequencies
$\omega_J^{(n)}$
come in two pairs $\pm \omega^{(n)}_{1,2}$, so that
for sufficiently large $n$ we have
\be
s_n = \omega^{(n)}_1+\omega^{(n)}_2 - 4n\;\;.
\ee
To estimate $s_n$, one can expand $\omega^{(n)}_1+\omega^{(n)}_2$ at large $n$.
For example, for $\hh=0$ one can use
\bea
(\omega^{(n)}_1+\omega^{(n)}_2)^2 &=&
\omega^{(n)2}_1 +\omega^{(n)2}_2 + 2(\omega^{(n)2}_1 \omega^{(n)2}_2)^{1/2}\non
&=&
4 \ff^2 + 8 n^2 - \kk_1 - \kk_2 +
((\kk_1-4n^2)(\kk_2-4n^2))^{1/2}\;\;.
\eea
Then one sees that the leading order ${\cal O}(n)$ contribution
cancels against
that of the fermions, and that the lower order terms are
\be
s_n = -{ 1 \ov 4  n } ( \kk_1 +  \kk_2 - 2\ff^2  )
 - { 1 \ov 64 n^3  }
 [  \kk_1^2 + \kk_2^2   - 2 \ff^2 (\kk_1 +  \kk_2) + 2 \ff^4 ]
   + {\cal O}( { 1 \ov n^5} )  \ .
\ee
The first term  here leads to a logarithmic divergence after summing over $n$:
$ \sum_{n=1}^\infty { 1 \ov n}e^{-\ep n}$ =$ - \ln \ep  + {\cal O}(\ep)$.
Its coefficient  $( 2 \ff^2 - \kk_1 - \kk_2)$
vanishes automatically  for the Ricci-flat spaces (cf. \rf{coo}, \rf{kop})
or otherwise is cancelled by the dilaton contribution \rf{dio},\rf{kop}
as in the similar dilatonic background case considered in \ci{prt}.
The resulting  expression for $\ep_0$ is thus finite,
in agreement  with  conformal invariance of the
theory before \lc\ gauge fixing.  This provides a useful check of the above
expressions for the frequencies.

For example, in the anti-Mach case \rf{anti} we find:
\be
s_n = - { \ff^4\ov 32 n^3}
 - { \ff^6\ov 128 n^5}  - { \ff^8\ov 8192 n^7} + {\cal O}( { 1 \ov n^9})
\ , \ee
and $\ep_0$ is automatically finite (and negative).
The negative vacuum energy  is analogous to the case of
\ci{prt} or \ci{zaff}; the sum can be evaluated using Epstein function as in
\ci{zaff}.

Likewise, when $\hh\neq 0$ but $\ff = 0$, one finds
\be
s_n = -\frac{1}{4 n}(\kk_1 + \kk_2 + 2\hh^2) + {\cal O}(\frac{1}{n^3})\;\;,
\ee
so that the logarithmic divergence in the sum \rf{noo}
is again absent when the conformal invariance condition
(\ref{coon}) is satisfied.

This fact, and  a somewhat surprising stronger statement, are true
for general values of $\ff,\hh$ and $\kk_i$. Even though the closed form
solution of (\ref{jok}) is unenlightning, we can find the explicit solution
of \rf{jok} in a series expansion in $1/n$. It is clear that the leading behaviour
of $\omega$ at large $n$ is its flat space value
$\omega^{(n)} = \pm 2n + {\cal O}(1)$. Then, expanding
\be
\omega^{(n)} = 2n + a + {b\ov n} + {c\ov n^2} + {\cal O}({1\ov n^3})\;\;,
\ee
one finds
\bea
a&=& \pm(\ff + \hh)\ , \non
b&=&\frac{1}{8}(2\ff^2 - 2\hh^2 - \kk_1 -\kk_2)\ ,  \non
c&=&\mp\frac{1}{16}\hh(2\ff^2 - 2\hh^2 - \kk_1 -\kk_2) \pm
\frac{(\kk_1-\kk_2)^2}{128(\ff+\hh)}\;\;.
\eea
If we now impose the conformal invariance condition  (\ref{coon}), i.e.
$2\ff^2 - 2\hh^2 = \kk_1 + \kk_2$, we obtain the simple expression
\be
\omega^{(n)} = 2n \pm (\ff+\hh) \pm \frac{(\kk_1-\kk_2)^2}{128 n^2(\ff+\hh)} + {\cal
O}(\frac{1}{n^3})\;\;.
\ee
Likewise, noting that (\ref{jok}) is invariant under $n \ra -n$ and $\hh\ra -\hh$,
the other two frequencies are
\be
\omega^{(n)} = -2n \pm (\ff-\hh) \pm \frac{(\kk_1-\kk_2)^2}{128 n^2(\ff-\hh)} + {\cal
O}(\frac{1}{n^3})\;\;.
\ee
We conclude that
the logarithmic divergence in the sum over $n$ is absent
for each series of  the frequencies $\omega_J^{(n)}$
independently  of  $J$.

%%%%%%%%%%%%%%%%%%%%%%%%%%%%%%%%%%%%%%%%%%%%%%%%%%%%%%%%%%%%%%%%%%%
\bigskip\bigskip
\bigskip\bigskip
%%%%%%%%%%%%%%%%%%%%%%%%%%%%%%%%%%%%%%%%%%%%%%%%%%%%%%%%%%%%%%%%%%%
\noindent
{\bf Acknowledgments}

We would like to thank  B. Pioline and J. Russo  for useful
discussions.  The research of M.B. is partially supported by EC contract
HPRN-CT-2000-00148.  The work of M.O. is supported in part by the European
Community's Human Potential Programme under contract HPRN-CT-2000-00131
Quantum Spacetime.  The research of  G.P. is partially supported by the
PPARC grant PPA/G/O/2000/00451 and by the European grant HPRN-2000-00122.
The work of A.A.T. is  supported in part by the grants INTAS  99-1590
and DOE DE-FG02-91ER40690, and by  the Royal Society Wolfson award.
  Both G.P. and A.A.T. are supported also by the PPARC SPG
grant PPA/G/S/1998/00613.

%%%%%%%%%%%%%%%%%%%%%%%%%%%%%%%%%%%%%%%%%%%%%%%%%%%%%%%

\newpage

\appendix

\section{Some facts about minors}
%%%%%%%%%%%%%%%%%%%%%%%%%%%%%%%%%%%%%%%%%%%%%%%%

\subsection{Null vectors and minors}

Let $M$ be a $(d\times d)$ matrix and let
$m_{ik}$ denote the minor of $M$, i.e.\ the determinant of the matrix one
obtains from $M$ by removing the $i$'th row and $k$'th column. These minors
satisfy the equations
\be
\sum_k M_{ik} m_{ik} (-1)^{i+k} = \det M
\label{id1}
\ee
for any $i$. Moreover, for any $M$ one has the identities
\be
\sum_k M_{lk} m_{ik} (-1)^k = 0 \;\;\;\;\mbox{for}\;\; l \neq i\;\;.
\label{id2}
\ee
This can be shown by noting that, whatever $M$ may be, the left-hand side
also calculates the determinant of a matrix with two equal rows, namely the determinant
of the matrix one obtains from $M$ by replacing $M_{ik}$ by $M_{lk}$. The determinant
of this new matrix is clearly zero and the identity follows.

In particular, therefore, when $\det M = 0$, one can choose the
null eigenvector $v_k$, $M_{ik}v_{k}=0$, to be
\be
v_k^{(i)} = (-1)^k m_{ik}
\ee
for any $i$. If there are no degeneracies, there should be a unique null
eigen-direction. Hence the vectors $v_k^{(i)}$ for different choices of $i$ should be
proportional to each other, i.e.
\be
\frac{v_k^{(i)}}{v_k^{(j)}} =\frac{v_l^{(i)}}{v_l^{(j)}}\;\;.
\ee
This can be rephrased as the useful identity
\be
m_{ik} m_{jl} = m_{il} m_{jk}\;\;,
\label{mmmm}
\ee
which is obviously true for $i=j$ or $k=l$ and
is indeed a property of matrices with $\det M = 0$ for $i\neq j, k\neq l$
because of the general identity
\be
m_{ik} m_{jl} - m_{il} m_{jk} =  \pm m_{ij,kl} \det M \;\;,
\ee
where $m_{ij,kl}$ is the secondary minor obtained by removing the two rows
and columns indicated. 

Another useful identity can be obtained by using the explicit form
of the matrix $M$. 
Let $\omega^{(n)}_J$ be one of the roots of $\det M=0$, where $M$ is
given in \rf{Mik}. Evaluating (\ref{id1}) at $\omega=\omega^{(n)}_J$
and summing over $i$, we get
\be
\sum_{i,k} M_{ik} m_{ik} (-1)^{k+i}=0~.
\ee
Substituting the expression for $M$ \rf{Mik}, we find
\bea
\sum_{i} ( (\omega^{(n)}_J)^2+k_i-4n^2) m_{ii}(\omega_J^{(n)})&+&
2i \omega_J^{(n)} \sum_{i,k} (-1)^{i+k} f_{ik} m_{ik}(\omega_J^{(n)})
\cr
&+&4in \sum_{i,k} (-1)^{i+k} h_{ik} m_{ik}(\omega_J^{(n)})=0~.
\label{idd1}
\eea

\subsection{Another identity for minors}

The general structure of the particle light-cone Hamiltonian
(before normal ordering) is
\be
H^{(0)} = \sum_{j,k=1}^d \sum_{\alpha,\beta = \pm}\xi_j^\alpha\xi_k^\beta
H^{\alpha\beta}(\omega_j,\omega_k)\ \ex{i(\alpha\omega_j + \beta\omega_k)\tau}
\ee
where
\be
H^{\alpha\beta}(\omega_j,\omega_k)= - \frac{1}{2}\sum_i
(\kk_i + \alpha\beta\omega_j\omega_k)m_{1i}(\alpha\omega_j)m_{1i}(\beta\omega_k)
\;\;.
\ee
Since in stationary coordinates the Hamiltonian is time-independent, it must be
true for instance that
\be
H^{++}(\omega_j,\omega_j)\sim
\sum_i (\omega_j^2 + \kk_i)m_{1i}(\omega_j)m_{1i}(\omega_j)=0\;\;\;\;\;\;\forall
\;\;j \;\;.
\ee
Noting that $(\omega_j^2 + \kk_i)$ is just the $(ii)$-component of $M(\omega_j)$,
we can also write this as
\be
\sum_{i=1}^d M_{ii}(\omega_j)m_{1i}(\omega_j)m_{1i}(\omega_j)=0\;\;\;\;\;\;\forall
\;\;j \;\;.\label{Ijj}
\ee
Here is a general proof of (\ref{Ijj}) which only makes use of the two elementary
identities (\ref{id1}),(\ref{id2}), which we will use in the form
\be
M_{11}m_{11}=\sum_{i=2}^d (-1)^i M_{1i}m_{1i}
\ee
and
\bea
&&\sum_{k=1}^d (-1)^k M_{ik} m_{1k} = 0 \;\;\;\;\;\;i\neq 1\non
\Ra && (-1)^1 M_{i1}m_{11} + (-1)^i M_{ii}m_{1i} = - \sum_{k\neq i,k\neq 1} (-1)^k
M_{ik} m_{1k}\;\;.
\eea
Then, using the antisymmetry of $M_{ik}$ for $i\neq k$, we can write
\bea
\sum_{i=1}^d M_{ii}m_{1i}m_{1i} &=& m_{11} \sum_{i=2}^d (-1)^i M_{1i}m_{1i}
+ \sum_{i=2}^d M_{ii}m_{1i}m_{1i}\non
&=&
\sum_{i=2}^d m_{1i}(M_{ii}m_{1i} - (-1)^i M_{i1}m_{11}) \non
&=&
\sum_{i=2}^d m_{1i}(-1)^i ((-1)^i M_{ii}m_{1i} - M_{i1}m_{11}) \non
&=&
-\sum_{i=2}^d \sum_{k\neq i,k=2}^m (-1)^{i+k}M_{ik}m_{1i}m_{1k}\non
&=& 0\;\;.
\eea

\section{The commutation relations of the $\xi$'s}
%%%%%%%%%%%%%%%%%%%%%%%%%%%%%%%%%%%%%%%%%%%%%%%%%%%%%%%%%%%%%%%%%%
\subsection{Imposing the canonical commutation relations}
%%%%%%%%%%%%%%%%%%%%%%%%%%%%%%%%%%%%%%%%%

Beginning with the string mode expansion (\ref{smz}),(\ref{smn}),
we now promote the $\xi$'s to operators and impose the CCRs
\bea
&& [X^i(\sigma,\tau),X^k(\sigma',\tau)]=
[\Pi^i(\sigma,\tau),\Pi^k(\sigma',\tau)]=0\non
&& [X^i(\sigma,\tau),\Pi^k(\sigma',\tau)]=i\d^{ik}\delta(\sigma-\sigma')\;\;,
\eea
where
\be
\delta(\sigma-\sigma')= \frac{1}{\pi}\sum_{n=-\infty}^{+\infty}
\ex{2in(\sigma-\sigma')}\;\;.
\ee
It will actually be sufficient to just impose the conditions
\bea
&& [X^i(\sigma,\tau),X^k(\sigma',\tau)]= 0\non
&& [\dot{X}^i(\sigma,\tau),X^k(\sigma',\tau)]=-i\pi\d^{ik}\delta(\sigma-\sigma')\;\;,
\eea
the rest following from the definition of the $\Pi^k$ in \rf{Pi}
and the equations of motion.

In terms of the diagonal commutators $C_j$ and $C_{J}^{(n)}$ defined in (\ref{cjcj}),
the CCRs become, for the zero mode,
\bea
&&\sum_{j=1}^d C_j
(m_{1i}(\omega_j)m_{k1}(\omega_j)-m_{1k}(\omega_j)m_{i1}(\omega_j))=0\non
&&\sum_{j=1}^d C_j \omega_j
(m_{1i}(\omega_j)m_{k1}(\omega_j)+m_{1k}(\omega_j)m_{i1}(\omega_j))
= \d_{ik}\;\;.
\eea
Now we can make use of the identity (\ref{mmmm}) to rewrite
\be
m_{1i}m_{k1}\pm m_{1k}m_{i1} = m_{11} (m_{ki}\pm m_{ik})\;\;.
\ee
Thus the conditions become
\bea
&&\sum_{j=1}^d C_j m_{11}(\omega_j)
(m_{ki}(\omega_j)-m_{ik}(\omega_j))=0\non
&&\sum_{j=1}^d C_j \omega_j m_{11}(\omega_j)
(m_{ki}(\omega_j)+m_{ik}(\omega_j))= \d_{ik}\;\;.
\label{Cj}
\eea
Similarly, for the string modes one obtains the conditions
\bea
&&\sum_{J=1}^{2d} C_J^{(n)} m_{11}(\omega_J^{(n)}) m_{ik}(\omega_J^{(n)})=0\non
&&\sum_{J=1}^{2d} C_J^{(n)} \omega_J^{(n)} m_{11}(\omega_J^{(n)})
m_{ik}(\omega_J^{(n)})= \d_{ik}\;\;.
\label{CJ}
\eea

%%%%%%%%%%%%%%%%%%%%%%%%%%%%%%%%%%
\subsection{Determining the $C_{J}$ using the Vandermonde and Lagrange polynomials}

We now explain how to determine the solutions to the equations
(\ref{Cj}),(\ref{CJ}) for the coefficients $C_j$ and $C_J^{(n)}$.
At first sight,  there appear to be too many equations for the $d$ ($2d$)
coefficients $C_j$ ($C_J$). However, a closer inspection of the minors
reveals that all of these are satisfied provided that these coefficients
satisfy $d$ ($2d$) conditions - thus the $C_j$ and $C_J$ are uniquely
determined by these equations.  The structure of the equations is
slightly different for the zero-modes and the string modes, so we will
discuss them separately, starting with the latter.

The key observation is that the highest power of $\omega$ that
can appear in any minor of a matrix $M$ of the form (\ref{Mik}) is
$\omega^{2d-2}$, arising from the product of $d-1$ diagonal entries. Such
a term can only arise for a diagonal minor $m_{kk}$. Furthermore,
the coefficient of $\omega^{2d-2}$ in such a minor is 1 for any one
of the $m_{kk}$. Moreover, $m_{kk}$ only arises in the commutator
$[\dot{X}^k,X^k]$. All other commutators will involve off-diagonal
minors and hence lower powers of $\omega$, $\omega^{i}$ with $i
=0,1,\ldots,2d-3$, and all these powers will appear. Hence all the $d^2$
equations (\ref{CJ}) are satisfied provided that the $2d$ coefficients
$C_J$ satisfy the $2d$ equations
\bea
&&\sum_{J=1}^{2d} C_J m_{11}(\omega_J)\omega_J^i = 0
\;\;\;\;\;\;\mbox{for}\;\;i=0,\ldots,2d-2\non
&&\sum_{J=1}^{2d} C_J m_{11}(\omega_J)\omega_J^{2d-1} = 1\;\;,
\eea
the extra power of $\omega_J$ in the last equation arising from the additional
$\omega_J$ in the second equation of (\ref{CJ}). These equations are solved by
\be
C_{J} =\frac{1}{m_{11}(\omega_J)\prod_{K\neq J}(\omega_J-\omega_K)}\;\;.
\label{CJs}
\ee
One way to see that this is the solution is to use the Vandermonde matrix
and its inverse, expressed in terms of Lagrange polynomials.

Let $x_i$, $i=0,\ldots,p-1$ be $p$ distinct
real or complex numbers. The Lagrange polynomials
$P_i(x)$ are defined by
\be
P_i(x) = \prod_{k\neq i} \frac{(x-x_k)}{(x_i-x_k)}\;\;.
\ee
They clearly satisfy
\be
P_i(x_k) = \d_{ik}\;\;.
\ee
Expanding the degree $(p-1)$ Lagrange polynomials as
\be
P_i(x) = \sum_{l=0}^{p-1}P_{il}x^l\;\;,
\ee
the above identity can be written as
\be
P_{il}x_k^l = \d_{ik}\;\;.
\ee
In other words, $P_{il}$ are the components of the matrix inverse to
the matrix $V$ with components
\be
V_{lk} = x_k^l\;\;.
\ee
This is just the Vandermonde matrix, and hence the coefficients of the Lagrange
polynomials give the inverse of the Vandermonde matrix. In particular,
\be
P_{i,p-1}= \frac{1}{\prod_{k\neq i} (x_i-x_k)}\;\;.
\ee
Thus, going back to our
problem of determining the $C_J$, the defining equations can be written as
\be
V_{LJ}C_Jm_{11}(\omega_J) = \d_{L,2d-1}\;\;,
\ee
and therefore the solution is
\be
C_J m_{11}(\omega_J)= P_{J,2d-1}
= \frac{1}{\prod_{K\neq J}(\omega_J-\omega_K)}\;\;,
\ee
as claimed.

For the zero-modes, the situation is slightly different.
 First of all, it follows
from the symmetry properties of the matrix $M$  that
in this case  $m_{ik}+m_{ki}$
has to be an even function of $\omega$, and $m_{ik}-m_{ki}$ an odd function of
$\omega$. Hence the equations (\ref{Cj}) only involve odd powers of $\omega$. Once
again, the highest power that can appear is $\omega_j^{2d-1}$ and it is these
terms that should give the non-vanishing term $\d_{ik}$ for $i=k$, whereas all
other odd powers should sum to zero to satisfy the remaining equations. Thus the
$d$ conditions on $d$ coefficients $C_j$ are
\bea
&&\sum_{j=1}^{d} C_j m_{11}(\omega_j)\omega_j^{2i-1} = 0
\;\;\;\;\;\;\mbox{for}\;\;i=1,\ldots,d-1\non
&&\sum_{j=1}^{d} C_j m_{11}(\omega_j)\omega_j^{2d-1} = \frac{1}{2}\;\;.
\eea
These equations are solved, in the same way as above, by
\be
C_{j} =
\frac{1}{2 m_{11}(\omega_j)\omega_j\prod_{k\neq j}(\omega_j^2-\omega_k^2)}\;\;.
\label{cjsol}
\ee

\section{Geometric Aspects of the Spectrum}
\subsection{The spectral hypersurface}

As we have seen,  the frequencies of the string are given
by the  vanishing condition \rf{detcon} of the determinant of the matrix $M$.
This condition can be viewed as defining a hypersurface $S$ in the space
of parameters and frequencies. For a generic choice
of parameters, the hypersurface $S$ is smooth.
Here we shall determine the values of the parameters
(and thus the corresponding plane-wave models)
which correspond
to  the singularities of this hypersurface.
The  singularities of
the hypersurface  are determined from the condition
that the first differential
of eq. \rf{detcon} with respect to the frequency $\omega$
and the parameters $k,f$ and $h$ vanishes on $S$.

We shall consider
string theory on the four-dimensional plane wave for which $M$ in
\rf{detcon} is given in \rf{mam}. In this case $S$ is a hypersurface
in $\bR^5$; we shall take the frequencies to be real but
the analysis can be easily extended  to  complex frequencies.
For $n=0$, we set $x=\omega^2$ and find
that the first differential of det$M$ is
\be
(2x + \kk_1 +\kk_2-4\ff^2) dx+(x+\kk_2) d\kk_1+(x+\kk_1) d\kk_2-8x\ff d\ff~.
\ee
The vanishing condition for this differential gives
\be
\kk_1=\kk_2\ , ~~~~~~\ff=0\ , ~~~~~~~~~~\omega^2=-\kk_1~.
\ee
Generically,  \rf{mam} has four distinct solutions for
 the frequency $\omega$.
The model at the singularity has only two distinct frequencies
 (for $\kk_1\not=0$).
Thus  the singularity occurs at a model with
degenerate frequencies. However,
the singularity does not describe all models with degenerate frequencies.
Next, let us take $n\not=0$;  in this case the hypersurface $S$ is singular
at
\bea
\omega^3+{1\over2} (\kk_1+\kk_2-4\ff^2-8n^2)\omega-4n\ff\hh&=&0
\cr
\ff\omega+2 n\hh=0\ , ~~~~~~~~\kk_1=\kk_2\ ,
~~~~~~~~~\omega^2&=&4n^2-\kk_1~ \ .
\eea
There is another equation which is implied by  the above.
There are two cases to consider. If $\ff=0$, then $\hh=0$ and the
frequencies are given by $\omega^2=4n^2-\kk_1$ ($\kk_1=\kk_2$).
If $\ff\not=0$, then
\be
\omega=-2 n \hh \ff^{-1}\ , ~~~~~~4n^2\hh^2+(\kk_1-4n^2)\ff^2=0\ .
\ee
Since for $n\not=0$ we expect  generically  four distinct frequencies, the
singularity occurs at a model with degenerate frequencies.

To describe all the models with degenerate frequencies by singularities,
we should consider
$S$ as a fibration over the space of parameters. For strings in four-dimensional
plane waves, $S$ should be thought of as a fibration over $\bR^4$ spanned by
the parameters $(\kk_1, \kk_2, \ff, \hh)$. The projection is the standard
projection of $\bR^5$ onto $\bR^4$ restricted on $S$. The generic fibre
of such a fibration has four points, the roots of the polynomial \rf{mam}.
It is clear now that $S$ as a fibration degenerates precisely when
two or more roots become equal. Thus at the singularities of this
fibration there are models with two or more degenerate frequencies.

There is a similar description for the hypersurface $S$ for strings
on all smooth homogeneous plane-waves. In this case $S$ is a fibration
over the space of parameters with fibre which has generically $2(d-2)$ points.
The fibration degenerates precisely when two or more frequencies become the same.

\subsection{Berry's Connection}

It is well known that during an adiabatic evolution along
a closed path in the space of parameters of a quantum mechanical
system, the wave functions  develop a phase
which  is the holonomy of the
Berry connection \cite{berry,simon}.  The quantum mechanical systems that
we are investigating in connection to string theory in smooth
homogeneous waves have such parameters. Adiabatic evolution
in the present context means that we allow the parameters,
like the rotation $f$ and the matrix $k$, to depend adiabatically on the
light-cone time $u$. The resulting quantum mechanical set-up becomes
precisely that which appears in the context of  Berry's connection.
We shall not investigate here the general case. Instead, we
shall focus on the quantum mechanical models that arise for strings in four-dimensional
plane-waves, like those described in section 2.6. In addition, we  shall assume
the generic case where there is no degeneracy of states.  We shall find that
the curvature of the Berry connection vanishes.

Using the unitary transformation  \rf{uni}, we can construct a basis in the Hilbert space of
the quantum mechanical  system described in section 2.6
as follows:
\be
|\psi_{n_1, n_2}>=U(\theta_1, \theta_2) |n_1, n_2>~,
\label{bbst}
\ee
where
\be
|n_1, n_2>={1\over \sqrt {n_1! n_2!}} (\aa_1^+)^{n_1} (\aa_2^+)^{n_2} |0>~
\ee
is the standard basis in the Hilbert space of a two-dimensional Harmonic
oscillator with frequencies $\omega_{1,2}$, masses $m_{1,2}$ and creation (annihilation)
operators $\aa^+_{1,2}$ ($\aa_{1,2}$).  We follow the notation of
section 2.6. Under the assumptions we have
made, these states are not degenerate and depend on the parameters
$\kk_1, \kk_2$ and $\ff$. The Berry connection is
\be
{\cal A}^{n_1, n_2}= <\psi_{n_1, n_2}| d |\psi_{n_1, n_2}>~,
\ee
where the exterior derivative $d$ is with respect to the parameters of the
system. Using \rf{bbst}, we can write
\be\la{bbb}
{\cal A}^{n_1, n_2}= <n_1, n_2| 
U^+(\theta_1, \theta_2) d U(\theta_1, \theta_2) |n_1, n_2>+
<n_1, n_2| d |n_1, n_2>~.
\ee
We shall show that the first term in the right-hand-side of \rf{bbb} vanishes.
Indeed, 
$$
<n_1, n_2| U^+(\theta_1, \theta_2) d U(\theta_1, 
\theta_2) |n_1, n_2> $$  $$=  \ -i d\theta_1
<n_1, n_2|U^+(\theta_1, \theta_2) xy U(\theta_1, \theta_2)|n_1, n_2>
- id\theta_2
<n_1, n_2| p_xp_y |n_1, n_2> $$ 
\be\la{bbc}
= -i d\theta_1
<n_1, n_2|U^+(\theta_1, \theta_2) xy U(\theta_1, \theta_2)|n_1, n_2>~.
\ee
However, as it was shown in \cite{yonei}, after adjusting for notation, we have
\bea
U^+(\theta_1, \theta_2) x U(\theta_1, \theta_2)&=&x+\theta_2 p_y
\cr
U^+(\theta_1, \theta_2) y U(\theta_1, \theta_2)&=&y+\theta_2 p_x~.
\eea
Thus we find
\bea
\rf{bbc}&=&-i \theta_2 d\theta_1  <n_1, n_2|(xp_x+p_y y) |n_1, n_2>
\cr
&=&
{1\over2} \theta_2 d\theta_1 <n_1, n_2|([\aa_1, \aa_1^+]- [\aa_2, \aa_2^+])|n_1, n_2>=0~.
\eea

It remains to commute the second term in the right-hand-side of \rf{bbb}. For this,
we first compute $<0|d|0>$ using
$$
<x,y|0>= N e^{-{1\over2} (m_1 \omega_1
x^2+ m_2 \omega_2 y^2)}
$$
and
$$
N=\left({m_1 m_2 \omega_1 \omega_2\over \pi^2}\right)^{1\over4}
$$
to find
\be
<0|d|0>= N^{-1} dN-{1\over4}  (m_1 \omega_1)^{-1} d (m_1 \omega_1)-{1\over4}
 (m_2 \omega_2)^{-1} d (m_2 \omega_2)=0~.
\ee
Next,  observe that
$$
d \aa^+_1= {1\over2} (m_1 \omega_1)^{-1} d(m_1 \omega_1) \aa_1~,~~~~~ d \aa^+_2
= {1\over2} (m_2 \omega_2)^{-1} d(m_2 \omega_2) \aa_2~.
$$
Using these relations and  also that 
$|n_1, n_2>=(n_1! n_2!)^{-{1\over2}} (\aa_1^+)^{n_1} (\aa^+_2)^{n_2} |0>$,
we finally find  $<n_1, n_2| d |n_1, n_2>=0$. Thus we conclude that
${\cal A}^{n_1, n_2}=0$ and so the Berry connection vanishes.
This may be due to the fact that the light cone
Hamiltonian is associated with a Klein-Gordon equation which is
real.  The same applies also to 
the quantum mechanical system associated with the string 
at level $n$ (for $\hh=0$).

%%%%%%%%%%%%%%%%%%%%%%%%%%%%%%%%%%%%%%%%%%%%%%%%%%%%%%%%%%%%%%%%%%%
\rnc{\Large}{\normalsize}

\end{document}